\documentclass[10pt,journal,compsoc]{IEEEtran}
\pdfoutput=1
\usepackage{graphicx}
\usepackage{color}
\usepackage[tight]{subfigure}
\usepackage{graphics}

\usepackage[sort, compress]{cite}
\usepackage{enumerate}
\usepackage{tikz}
\usetikzlibrary{shapes,arrows,positioning}

\usepackage{bbm}
\usepackage{amssymb}
\usepackage{amsmath}
\usepackage{amsfonts}
\usepackage{latexsym}
\usepackage{enumitem}
\usepackage{verbatim}

\usepackage{textcomp}
\usepackage[latin1]{inputenc}

\usepackage{multirow}

\usepackage{booktabs}

\usepackage{url}

\usepackage{balance}

\usepackage{caption}
\newcommand{\code}[1]{\texttt{#1}}
\usepackage[lined,boxed,linesnumbered,vlined]{algorithm2e}
\usepackage{algpseudocode}

\usepackage{etoolbox}

\def\equationautorefname~#1\null{%
  (#1)\null
}

\SetKwInput{Problem}{problem}
\SetKwInput{Input}{input}
\SetKwInput{Output}{output}
\SetAlgoSkip{smallskip} 
\SetAlgoInsideSkip{smallskip} %
\usepackage{amsthm}
\newtheoremstyle{saber}
  {}
  {}
  {}
  {}
  {\bfseries}
  {.}
  {.5em}
  {}

\theoremstyle{saber}
\ifcsundef{theorem}{
\newtheorem{theorem}{Theorem}
}{}
\ifcsundef{corollary}{%
}{}
\ifcsundef{lemma}{%
\newtheorem{lemma}{Lemma}
}{}
\ifcsundef{definition}{%
\newtheorem{definition}{Definition}
}{}
\ifcsundef{fact}{%
}{}
\ifcsundef{remark}{%
}{}

\ifx\theoremautorefname\undefined
\newcommand*{\theoremautorefname}{Theorem}
\fi
\ifx\lemmaautorefname\undefined
\newcommand*{\lemmaautorefname}{Lemma}
\fi
\ifx\definitionautorefname\undefined
\newcommand*{\definitionautorefname}{Definition}
\fi
\ifx\corollaryautorefname\undefined
\newcommand*{\corollaryautorefname}{Corollary}
\fi
\ifx\factautorefname\undefined
\newcommand*{\factautorefname}{Fact}
\fi

\ifx\argmax\undefined
\newcommand{\argmax}{\operatornamewithlimits{argmax}}
\fi
\ifx\argmin\undefined
\newcommand{\argmin}{\operatornamewithlimits{argmin}}
\fi
\ifx\limsup\undefined
\newcommand{\limsup}{\operatornamewithlimits{limsup}}
\fi
\ifx\liminf\undefined
\newcommand{\liminf}{\operatornamewithlimits{liminf}}
\fi
\ifx\norm\undefined
\newcommand{\norm}[1]{\lVert#1\rVert}
\fi
\ifx\abs\undefined
\newcommand{\abs}[1]{\lvert#1\rvert}
\fi
\ifx\set\undefined
\newcommand{\set}[1]{\left\{#1\right\}}
\fi
\ifx\mset\undefined
\newcommand{\mset}[1]{\lbrack #1\rbrack}
\fi
\ifx\etal\undefined
\newcommand{\etal}[1]{{\em #1 et al.}~}
\fi
\ifx\ie\undefined
\newcommand{\ie}{{\em i.e.,} }
\fi
\ifx\eg\undefined
\newcommand{\eg}{{\em e.g.,} }
\fi
\ifx\etc\undefined
\newcommand{\etc}{{\em etc.,} }
\fi
\ifx\wrt\undefined
\newcommand{\wrt}{{\em w.r.t.} }
\fi
\ifx\dotprod\undefined
\newcommand{\dotprod}[2]{
  \langle #1, #2 \rangle
}
\fi
\ifx\iid\undefined 
\newcommand{\iid}{i.i.d.}
\fi
\ifx\bigParenthes\undefined 
\newcommand{\bigParenthes}[1]{
  \big(#1\big)
}
\fi
\ifx\bigBracket\undefined 
\newcommand{\bigBracket}[1]{
  \big\{#1\big\}
}
\fi
\ifx\bigSqBracket\undefined 
\newcommand{\bigSqBracket}[1]{
  \big[#1\big]
}
\fi
\ifx\BigParenthes\undefined 
\newcommand{\BigParenthes}[1]{
  \Big(#1\Big)
}
\fi
\ifx\BigBracket\undefined 
\newcommand{\BigBracket}[1]{
  \Big\{#1\Big\}
}
\fi
\ifx\BigSqBracket\undefined 
\newcommand{\BigSqBracket}[1]{
  \Big[#1\Big]
}
\fi
\ifx\biggParenthes\undefined 
\newcommand{\biggParenthes}[1]{
  \bigg(#1\bigg)
}
\fi
\ifx\biggBracket\undefined 
\newcommand{\biggBracket}[1]{
  \bigg\{#1\bigg\}
}
\fi
\ifx\biggSqBracket\undefined 
\newcommand{\biggSqBracket}[1]{
  \bigg[#1\bigg]
}
\fi
\ifx\BiggParenthes\undefined 
\newcommand{\BiggParenthes}[1]{
  \Bigg(#1\Bigg)
}
\fi
\ifx\BiggBracket\undefined 
\newcommand{\BiggBracket}[1]{
  \Bigg\{#1\Bigg\}
}
\fi
\ifx\BiggSqBracket\undefined 
\newcommand{\BiggSqBracket}[1]{
  \Bigg[#1\Bigg]
}
\fi
\ifx\bracket\undefined
\newcommand{\bracket}[1]{
  \{#1\}
}
\fi
\ifx\parenthes\undefined
\newcommand{\parenthes}[1]{
  (#1)
}
\fi
\ifx\sqBracket\undefined
\newcommand{\sqBracket}[1]{
  [#1]
}
\fi
\ifx\prob\undefined 
\newcommand{\prob}[1]{\mathbb{P}[#1]}
\fi
\ifx\Prob\undefined 
\newcommand{\Prob}[1]{\mathbb{P}\big[#1\big]}
\fi
\ifx\probb\undefined 

\fi
\ifx\expect\undefined 
\newcommand{\expect}[1]{\mathbb{E}[#1]}
\fi
\ifx\Expect\undefined 
\newcommand{\Expect}[1]{\mathbb{E}\big[#1\big]}
\fi
\ifx\expectt\undefined 
\newcommand{\expectt}[1]{\mathbb{E}\bigg[#1\bigg]}
\fi
\ifx\walk\undefined 
\makeatletter
\newcommand{\walk}[1]{%
  \@tempswafalse
  \@for\next:=#1\do
    {\if@tempswa\!\!\rightarrow\!\!\else\@tempswatrue\fi\next}%
}
\makeatother
\fi
\ifx\seq\undefined 
\newcommand{\seq}{\!=\!}
\fi
\ifx\sminus\undefined 
\newcommand{\sminus}{\!-\!}
\fi
\ifx\sm\undefined 
\newcommand{\sm}[1]{\!#1\!}
\fi

\ifx\union\undefined
\newcommand{\union}[2]{#1\!\cup\!#2}
\fi

\ifx\hl\undefined 
\newcommand{\hl}[2]{{\color{#1}#2}}
\fi

\ifx\hlc\undefined 
\newcommand{\hlc}[3]{{\color{#1}#2 [remark: #3]}}
\fi

\ifcsdef{sitemize}{}{

}

\ifcsdef{senumerate}{}{

}

\ifcsdef{spenumerate}{}{
\newenvironment{spenumerate}{%
   \begin{list}{(\arabic{enumi})}{%
    \setlength\labelwidth{3.5em}%
    \setlength\leftmargin{2.5em}%
    \setlength{\topsep}{2pt plus 2pt minus 2pt}%
    \setlength\itemsep{0.0cm}%
    \usecounter{enumi}}%
  }{\end{list}}
}

\ifcsdef{slenumerate}{}{

}

\makeatletter
\newcommand*\bigdot{\mathpalette\bigdot@{.5}}
\newcommand*\bigdot@[2]{\mathbin{\vcenter{\hbox{\scalebox{#2}{$\m@th#1\bullet$}}}}}
\makeatother

\def\calL{ \mathcal{L} }

\usepackage{balance}

\def \IQExample {
    \includegraphics[width=2.8in]{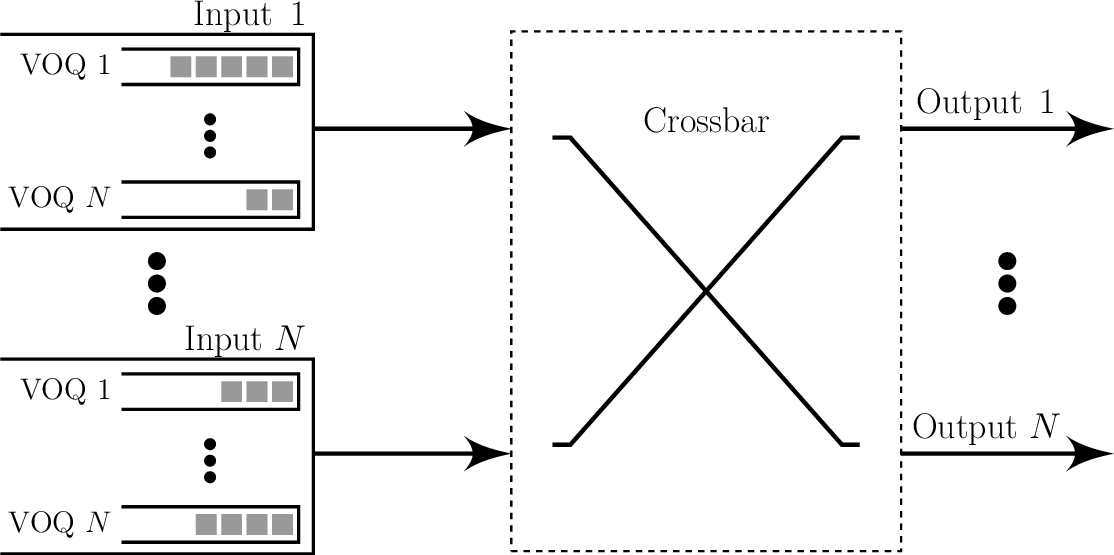}
}

\def \MeanDelayvsLoadsO {
    \includegraphics[width=0.90\textwidth]{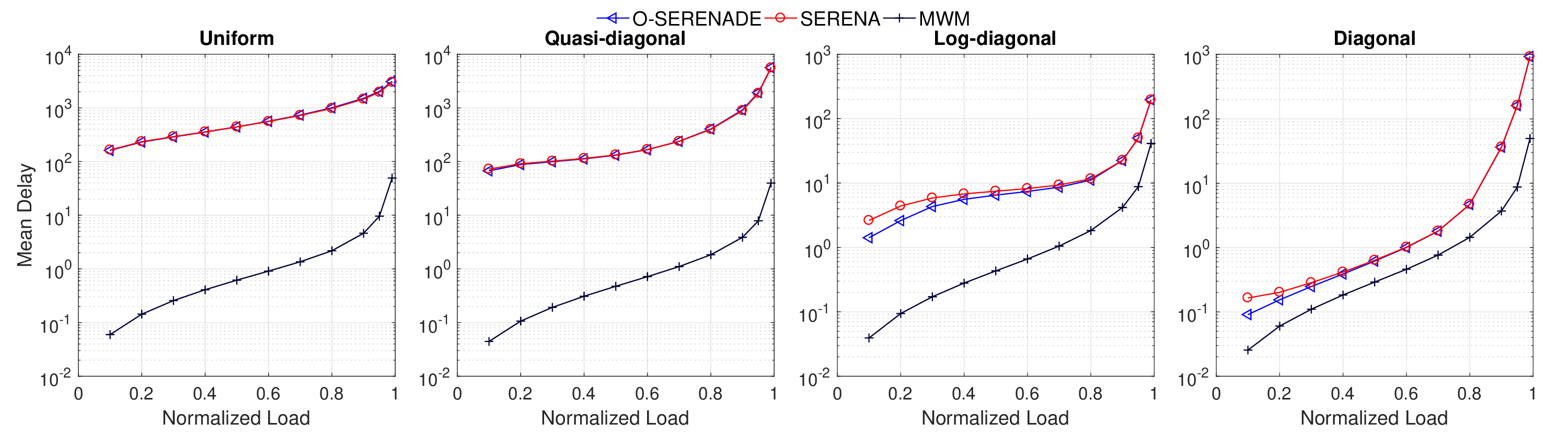}
}

\def \MeanDelayvsBurstSize {
     \includegraphics[width=0.90\textwidth]{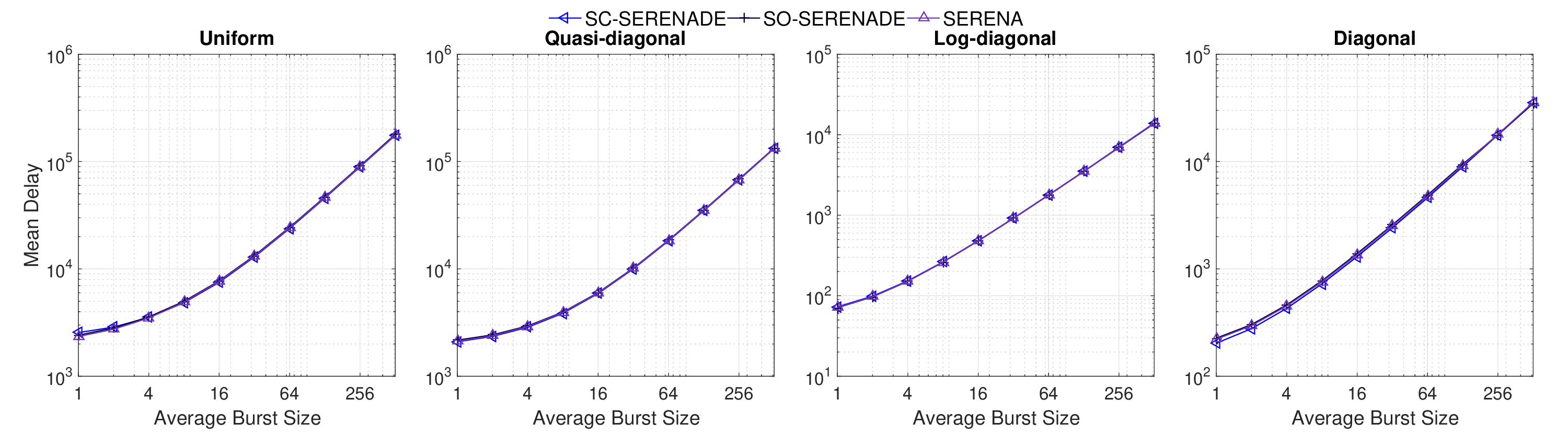}
}

\def \MeanDelayvsBurstSizeModerate {
     \includegraphics[width=0.90\textwidth]{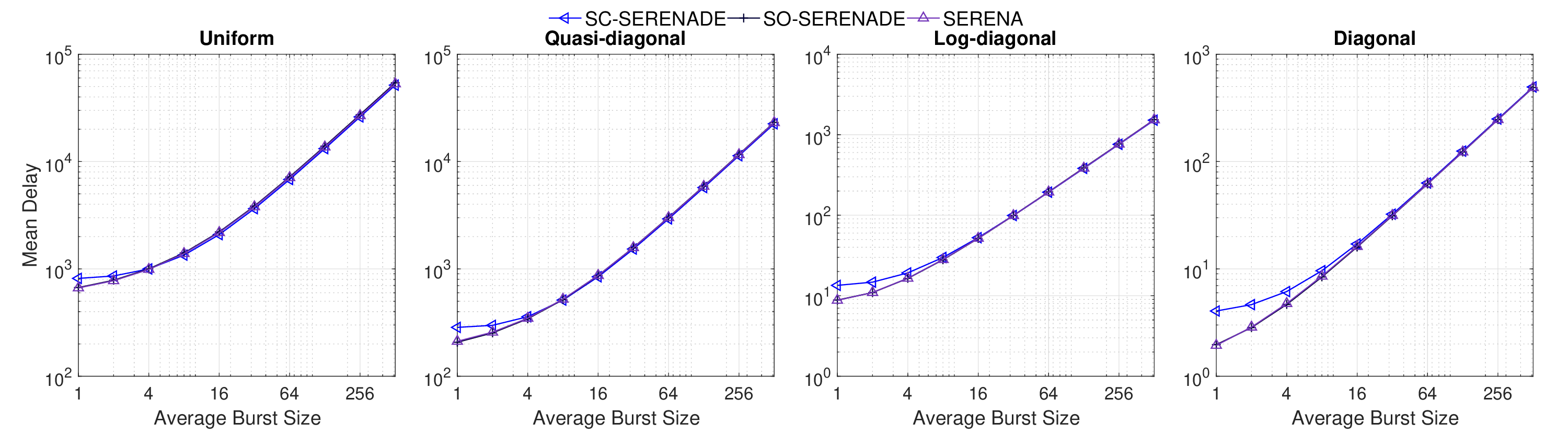}
}

\newcommand{\sleader}{
    \mathcal{L}
}

\usepackage{./libs/saber_revision}
\Extensionfalse
\Highlightfalse
\usepackage{etoolbox}

\patchcmd{\IEEEproofindentspace}{2\parindent}{0pt}{}{}

\begin{document}

\title{SERENADE: A Parallel Randomized Algorithm for Crossbar Scheduling in Input-Queued Switches}





%
\author{Long~Gong,
Liang~Liu,
Sen~Yang,
Jun~(Jim)~Xu,
Yi~Xie,
Xinbing~Wang
\thanks{L. Gong, L. Liu, S. Yang, and J. Xu are with Georgia Institute of Technology, Atlanta,
GA, 30332 (e-mail: gonglong@gatech.edu, lliu315@gatech.edu, sen.yang@gatech.edu, and jx@cc.gatech.edu).}
\thanks{Y. Xie is with Xiamen University, China (email: csyxie@xmu.edu.cn).}
\thanks{X. Wang is with Shanghai Jiao Tong University, China (email: xwang8@sjtu.edu.cn).}
\thanks{Manuscript received March 19, 2019.}
}

\IEEEtitleabstractindextext{%
\begin{abstract}
Most of today's high-speed switches and routers adopt an
input-queued crossbar switch architecture.  Such a switch needs to 
compute a matching (crossbar schedule) 
between the input ports and output ports 
during each switching cycle (time slot).  A key research challenge
in designing large (in number of input/output ports $N$) input-queued crossbar switches is to develop 
crossbar scheduling algorithms that can
compute ``high quality" matchings -- \ie those that result in
high switch throughput (ideally $100\%$) and low queueing
delays for packets -- at line rates. 
\modifiedHL{
SERENA is one such algorithm: 
it outputs excellent matching decisions that result in $100\%$ switch throughput 
and reasonably good queueing delays.
} 
However, since SERENA is a centralized algorithm with $O(N)$ computational complexity, it cannot support switches that both are  
large 
and have a very high line rate per port.  
\modifiedHL{
In this work, we propose 
SERENADE (SERENA, the Distributed Edition), a parallel iterative algorithm 
that emulates SERENA 
in only $O(\log N)$ 
iterations between input ports and output ports, and hence has a time complexity
of only $O(\log N)$ per port. 
We prove that SERENADE can exactly emulate SERENA. We also propose an early-stop version of 
SERENADE, called O-SERENADE, to only approximately emulate SERENA. Through extensive simulations, we 
show that O-SERENADE can achieve 
100\% throughput and that it has similar as or slightly better delay performance than 
SERENA under various load conditions and traffic patterns.
}
\end{abstract}

} 

\maketitle

%
\IEEEpeerreviewmaketitle


\section{Introduction}
\label{sec: introduction}



The volumes of network traffic across the Internet and in data-centers continue to grow relentlessly,
thanks to existing and emerging data-intensive applications, such as Big Data analytics, cloud computing,
and video streaming.  At the same time, the number of network-connected devices also grows explosively,
fueled by the wide adoption of 
smart phones and the emergence of the Internet of things.
%
\modifiedHL{
To transport and ``direct'' this massive amount of traffic to their respective destinations, 
switches and routers capable of connecting a large number of ports 
(called high-radix), and operating at very high per-port speeds are badly needed. 
Motivated by this rapidly growing need, much research has been devoted to, and significant breakthroughs
made, on the hardware designs for fast high-radix switching fabrics recently ({\it e.g.},~\cite{cakir2016HighRadix,Dai2017HighRadix}). Indeed,
these recent breakthroughs have made fast high-radix switching hardware not only technologically feasible
but also economically and environmentally (more energy-efficient) favorable, as compared to low-radix
switching hardware~\cite{cakir2016HighRadix}.
}

\begin{figure}
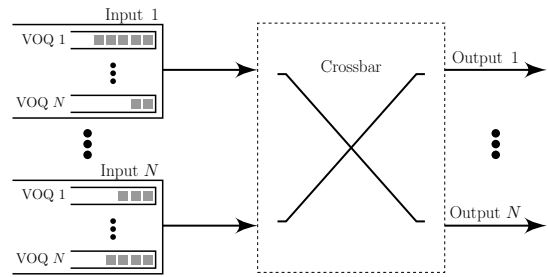

\centering
\IQExample
\caption{An input-queued crossbar switch.}
\label{fig: input-queued-switch}
\end{figure}

Most of today's switches and routers adopt an input-queued crossbar switch architecture.
\autoref{fig: input-queued-switch} shows a generic input-queued
switch employing a crossbar to interconnect $N$ input ports with $N$ output
ports.  Each input port has $N$ Virtual Output Queues
(VOQs). A VOQ $j$ at input port $i$ serves as a buffer for the packets
going into input port $i$ destined for output port $j$. The use of VOQs solves the Head-of-Line
(HOL) blocking issue~\cite{KarolHluchyjMorgan1987HOL}, which
severely limits the throughput of the
switching system.

\subsection{Matching Problem and SERENA Algorithm}


In an input-queued crossbar switch, each input port can be connected to only one output
port, and vice versa, in each switching cycle, or time slot.
Hence, it needs to compute, per time slot,
a one-to-one \textit{matching} between input and output ports. 
\modifiedHL{
One major research challenge
in designing high-radix input-queued crossbar switches is to develop algorithms that can
compute ``high quality" matchings 
}
-- \ie those that result in
high switch throughput (ideally $100\%$) and low queueing
delays for packets -- at line rates.

Unfortunately, there appears to be a tradeoff
between the quality of a matching and the amount of 
time needed to compute it (\ie time complexity). 
\modifiedHL{
For example, when used for crossbar scheduling,
Maximum Weight Matching (MWM), with a suitable weight measure, 
is known to result in 100\% switch throughput~\cite{McKeownMekkittikulAnantharamEtAl1999} and 
is conjectured to have the optimal delay performance~\cite{ShahWischik2006MWM0}. 
}
Each matching decision however takes $O(N^3)$ time
to compute~\cite{EdmondsKarp1972}.

Ideally, a crossbar scheduling algorithm should have time complexity much lower
than $O(N^3)$, but performance close to that of MWM.
SERENA~\cite{ShahGiacconePrabhakar2002SerenaMicro,GiacconePrabhakarShah2003SerenaJSAC} is one such algorithm. 
SERENA produces excellent matching decisions that result in $100\%$ switch throughput
and 
{\bf reasonably good queueing delay}. 
However, it is a centralized algorithm with $O(N)$ time complexity.  When $N$ is large,
this complexity is too high to support very high link rates.  Hence, as stated in~\cite
{ShahGiacconePrabhakar2002SerenaMicro,GiacconePrabhakarShah2003SerenaJSAC},
SERENA is designed for high-aggregate-rate switches -- \ie those that have either a large number of ports 
or a
very high line rate per port -- but not for those that have both.
While parallelizing the SERENA algorithm seems to be an obvious solution to this scalability problem,
we will show in \autoref{subsec: merge} that a key procedure in SERENA, namely MERGE,
is monolithic in nature, making SERENA hard to parallelize.

\deletedHL{
While considerable research has been performed on crossbar scheduling algorithms 
for decades, these algorithms~\cite{Anderson1993PIM,McKeown1995iLQF,Tassiulas1998TASS,Mekkittikul98stability,McKeownMekkittikulAnantharamEtAl1999,McKeown99iSLIP,Goudreau00Shakeup,ShahGiacconePrabhakar2002SerenaMicro,Giaccone2002Apsara,GiacconePrabhakarShah2003SerenaJSAC,Aggarwal2003EdgeColoring,Keslassy2003MSM,ZhangBhuyan2003iDRR,Fayyazi04ParallelMWM,ShahWischik2006MWM0,BayatiPrabhakarShahEtAl2007Iterative,Neely2007FrameBased,Gupta09Node,Ye10Variable,AtallaCudaGiacconeEtAl2013BPAssist,Wang2018ParallelEdgeColoring,Hu2018HRF} are not suitable 
for future high-speed large IQ switches, since what was 
considered high-speed ({\it e.g.,} $1$ Gbps per port) and large ({\it e.g.,} 
$64$ ports) is pale in comparison with the type of speed ({\it e.g.,} $1$ Tbps per 
port) and large ({\it e.g.,} $1,000$ ports) we are aiming for today~\cite{DeCusatis2014FutureSwitch,Kim2005RouterDevTrand}, yet 
our expectation of delay and throughput performance are only going to become 
higher. For example, the two well-known among these algorithms are SERENA~\cite{ShahGiacconePrabhakar2002SerenaMicro,GiacconePrabhakarShah2003SerenaJSAC} and 
iSLIP~\cite{McKeown99iSLIP}, but neither is suitable for 
high-speed IQ switches with large number of ports. On one hand, although SERENA 
can achieve 100\% throughput and has excellent delay performance close to 
that of MWM, its computational 
complexity for generating a matching is $O(N)$, which is too high for high-speed 
switches when $N$ is large. On the other hand, 
iSLIP only requires $O(\log N)$ 
iterations to compute a matching, but it can only guarantee to achieve at least 
$50\%$ throughput~\cite{Dai00Speedup} and it suffers from poor delay performance under heavy nonuniform 
traffic~\cite{GiacconePrabhakarShah2003SerenaJSAC}.}
\deletedHL{
Researchers have been searching for crossbar scheduling algorithms that have time complexity much lower
than $O(N^3)$, but performance
close to MWM. 
Some researchers have developed parallel or distributed MWM algorithms\cite{Fayyazi04ParallelMWM,Fayyazi2006ParaMWM,BayatiPrabhakarShahEtAl2007Iterative,BayatiShahSharma2008MWMMaxProduct} that, by 
distributing the computational cost across multiple processors (nodes), bring down 
the per-node computational complexity. However, these algorithms either require too many 
processors ~\cite{Fayyazi04ParallelMWM,Fayyazi2006ParaMWM} or have very high per-node computational complexity 
~\cite{BayatiPrabhakarShahEtAl2007Iterative,BayatiShahSharma2008MWMMaxProduct}, which are impractical to implement in today's IQ switches. 

Some researchers have proposed algorithms based on maximum 
cardinality matching or maximal matching, 
such as PIM~\cite{Anderson1993PIM} and iSLIP~\cite{McKeown99iSLIP}. 
Those types of matchings can be computed with lower computational complexity, they 
however have lower qualities than MWM: They can not achieve 100\% switch throughput and 
usually have much higher delays than MWM under non-uniform traffic. 

Some researchers have proposed  
low-complexity randomized algorithms~\cite{Tassiulas1998TASS,ShahGiacconePrabhakar2002SerenaMicro,GiacconePrabhakarShah2003SerenaJSAC,Ye10Variable}. 
Tassiulas~\cite{Tassiulas1998TASS} proposed the so-called ``pick-and-compare'' algorithm, which can achieve $100\%$ switch 
throughput with linear complexity but has very bad delay performance. Inspired by 
Tassiulas' work, 
Giaccone, Prabhakar, and Shah proposed 
SERENA~\cite
{ShahGiacconePrabhakar2002SerenaMicro,GiacconePrabhakarShah2003SerenaJSAC}.
SERENA produces excellent matching decisions that result in $100\%$ switch throughput
and queueing delay close to that of MWM.
However, it is a centralized algorithm with $O(N)$ time complexity.  When $N$ is large,
this complexity is too high to support very high link rates.  Hence, as stated in~\cite
{ShahGiacconePrabhakar2002SerenaMicro,GiacconePrabhakarShah2003SerenaJSAC},
SERENA is designed for high-aggregate-rate switches -- \ie those that have either a large number of ports 
or a
very high line rate per port -- but not for those that have both. } 
\deletedHL{
While parallelizing the SERENA algorithm seems to be an obvious solution to this scalability problem,
we will show in \autoref{subsec: merge} that a key procedure in SERENA, namely MERGE,
is monolithic in nature, making SERENA hard to parallelize.}

\deletedHL{
Besides the above algorithms which make a matching decision per each time slot, some 
researchers have proposed another type of crossbar scheduling -- frame-based 
scheduling~\cite{Aggarwal2003EdgeColoring,Neely2007FrameBased,Wang2018ParallelEdgeColoring}. In frame-based scheduling, consecutive time slots are grouped 
into frames and the matching decisions are made on a per-frame basis. Since a frame 
typically spans many time slots ({\it e.g.,} $O(\log N)$), and a packet arriving at 
the beginning of a frame has to wait till the end of the frame, frame-based 
scheduling typically results in higher queueing delays.      
}
%
%
%
\subsection{Parallelizing SERENA via SERENADE}\label{subsec: intro-parallelize-serena}
\modifiedHL{
In this work, we propose SERENADE (SERENA, the Distributed Edition), 
a parallel iterative algorithm 
that emulates each
matching computation of SERENA
using only $O(\log N)$ 
iterations between input ports and output ports.
}
Hence, each input or output port
needs to do only $O(\log N)$ work to compute a matching, making SERENADE scalable in both the switch size and
the line rate per port. 

\modifiedHL{
SERENADE consists of two stages: knowledge-discovery stage and distributed binary search stage. 
The knowledge-discovery stage uses a knowledge-discovery 
procedure, which has at most $1\!+\!\log_2N$ iterations, to gather information at each input port. 
After this stage, some input ports might be able to make the same decisions as they would under SERENA, 
whereas other input ports are not able to do so. 
Then, in the distributed binary search stage, those input ports will also be  
able to make the same matching decisions as they would do under SERENA 
by performing an additional distributed 
binary search, which has at most 
$\log_2N$ iterations, guided by the information 
gathered during the knowledge-discovery stage. We prove that 
SERENADE exactly emulates SERENA. 
}

\deletedHL{
Parallelizing existing algorithms is an obvious and commonly-used solution 
to lowering the per-node 
computational complexity, SERENA is however hard to parallelize because of 
the monolithic nature of MERGE, a key procedure in SERENA, which we will show in \autoref{subsec: merge}.}

SERENADE overcomes 
the aforementioned challenge of parallelizing SERENA, namely the monolithic nature of the MERGE procedure,
by making do with less. 
More specifically, we will show 
toward the end of~\autoref{subsubsec:knowledge-sets}, 
\modifiedHL{
in SERENADE,
}
after its $O(\log N)$ iterations,
each input port has much less information to work with than the (central) switch controller in SERENA. 
In other words, SERENADE does not precisely
parallelize SERENA, in that it does not duplicate the full information gathering capability of SERENA;  rather, it gathers
just enough information needed to make a matching decision that is either exactly or almost as wise.  This making do with less is a
major innovation and contribution of this work. 

\modifiedHL{
To reduce the complexities, we propose an early-stop version of SERENADE, called 
O-SERENADE, to approximately emulate SERENA. O-SERENADE gets rid of the distributed binary search and 
only approximately emulates SERENA by making opportunistic matching 
decisions after the knowledge-discovery stage. 
Despite this approximation, the delay performance of O-SERENADE is similar to or slightly better than that of SERENA, under 
various load conditions and traffic patterns.
}

\deletedHL{
SERENADE, by ``converting" the sequential algorithm SERENA that has $O(N)$ time complexity to a parallel iterative algorithm
that requires only $O(\log N)$ iterations, also has the following profound implication.
By far 
one 
well-known iterative crossbar scheduling algorithm that achieves great commercial success 
is iSLIP~\cite{McKeown99iSLIP}.
As mentioned above, 
iSLIP only requires $O(\log N)$ iterations, it however computes 
maximal matchings.  
Hence, iSLIP cannot achieve 100\% throughput
except under the uniform traffic pattern, and has much longer
queueing delays than SERENA under heavy nonuniform traffic~\cite{GiacconePrabhakarShah2003SerenaJSAC}.  SERENADE gets the better of both worlds:  Its time and communication complexities
are comparable to iSLIP's, yet its throughput and delay performances are either identical or close to SERENA's. 
To the best of our knowledge, SERENADE provides the first logarithm-time parallel 
iterative algorithm, requiring only $N$ processors, 
which can achieve 100\% throughput and have 
good delay performance close to MWM's. We hope that the approaches we used in 
SERENADE can inspire parallelizations of other good centralized algorithms.}


The rest of the paper is organized as
follows.  In~\autoref{sec: background}, we offer some background on 
input-queued crossbar switches. 
In~\autoref{sec: serena-and-merge}, we describe in detail the MERGE procedure in SERENA
that is to be parallelized in SERENADE.
\modifiedHL{
In~\autoref{sec: serenade}, we provide an overview of SERENADE, before zooming in on the knowledge-discovery 
stage in~\autoref{sec:knowledge-discovery}, its augmentation, {\it i.e.,} the leader election in~\autoref{sec:leader-election}, and the distributed binary search stage in~\autoref{sec:bs}. 
In~\autoref{subsec: o-serenade}, we introduce the early-stop version, O-SERENADE. 
}
In~\autoref{sec:evaluation}, we evaluate the performance of SERENADE. 
In~\autoref{sec: related-work}, we provide a brief survey of related work before concluding the paper in~\autoref{sec: conclusion}.
\section{System Model and Background}
\label{sec: background}








We assume that
all incoming
variable-size packets are segmented into fixed-size packets,
which are then reassembled when leaving the switch.
Hence we consider the switching of only fixed-size packets in the sequel, and each
such fixed-size packet takes exactly one time slot to transmit.
\modifiedHL{
We also assume that both the output ports and  
the crossbar operate at the same speed as the input ports. 
}
Both assumptions above are widely adopted in the 
literature~\cite {Mekkittikul98stability,McKeownMekkittikulAnantharamEtAl1999,McKeown99iSLIP,GiacconePrabhakarShah2003SerenaJSAC,Gupta09Node}. 
\modifiedHL{
Like in~\cite{GiacconePrabhakarShah2003SerenaJSAC}, we further assume that the arrival processes for those 
fixed-size packets are {\it i.i.d.} Bernoulli. 
}

An $N \!\times\! N$ input-queued crossbar switch is usually modeled as a weighted complete bipartite graph $G(I\bigcup O)$, with
the $N$ input ports and the $N$ output ports represented as the two disjoint vertex sets $I \!=\! \{I_1, I_2, \dots, I_N\}$ and $O \!=\! \{O_1, O_2, \dots, O_N\}$ respectively.
Each edge
$(I_i, O_j)$ corresponds to the $j^{th}$ VOQ at input port $i$ and its weight is defined as the number of packets in the VOQ. 
We denote this edge also as $I_i\!\rightarrow\!O_j$ when its direction is emphasized. 

A valid schedule, or
\textit{matching}, is a set of edges between $I$ and $O$
in which no two distinct edges share a vertex.
The weight of a matching is defined as the total weight of all edges belonging to the matching.
We say that a matching is {\it full} if all vertices in $G(I\bigcup O)$ are an endpoint of an edge in the matching, and is {\it partial} otherwise.
Clearly, in an $N \!\times\! N$ switch, any full matching contains exactly $N$ edges.

\section{SERENA}
\label{sec: serena-and-merge}


\deletedHL{
SERENA is an adaptive algorithm designed to eventually converge to MWM in the sense that 
the weight of the matching, at any future time slot after the convergence period, is with high probability either equal to or close to that of MWM.}
\deletedHL{
This adaptive algorithm}

To explain SERENADE, we first explain SERENA, the algorithm it parallelizes.  
SERENA 
consists of two steps.  The first step is to derive a full matching $R(t)$ from the set of packet arrivals $A(t)$.
The second step is to {\it merge} $R(t)$ with the full matching $S(t-1)$ used in the previous time slot, to arrive at the full matching $S(t)$ to be used for 
the current time slot $t$.  
\modifiedHL{
After we briefly describe the first step, 
}
we will focus on the second step, MERGE, in the rest of this section. 

In~\cite{GiacconePrabhakarShah2003SerenaJSAC}, the set of packet arrivals $A(t)$ is modeled as an {\it arrival graph}, which we denote
also as $A(t)$, as follows:  an edge $(I_i, O_j)$ belongs to $A(t)$ if and only if there is a packet arrival\footnote{\modifiedHL{According to the {\it i.i.d.} Bernoulli assumption in~\autoref{sec: background}, there is at most one packet arrival to any input port during each time slot.}}
to the corresponding VOQ at time slot $t$.  
Note that $A(t)$ is not necessarily a matching, 
because more than one input ports could have 
a packet arrival (\ie edge) destined for the same output port at time slot $t$.  
Hence in this case, each output port prunes all such edges incident upon it except the one with the heaviest weight 
(with ties broken randomly). The pruned graph, denoted as 
$A'(t)$, is now a matching. 

This matching $A'(t)$, which is typically partial, is then randomly populated 
into a full matching $R(t)$ by pairing the yet unmatched input ports with the yet unmatched output ports in a round-robin manner.
Although this POPULATE procedure alone, with the round-robin pairing, 
has $O(N)$ computational complexity, we 
will show in~\autoref{subsec: parallel-population} 
that it can be reduced to the computation of prefix sums and solved using the
classical parallel algorithm~\cite{Ladner1980PrefixSum} whose time complexity in this context is $O(\log N)$ per port.  
%
%
%



\subsection{Overview of The MERGE Procedure}
\label{subsec: merge}


\deletedHL{
This gradual increase of weight over time allows the matching $S(t)$ to converge towards MWM as $t$ increases.}



In this section, 
\modifiedHL{we explain the MERGE procedure through which SERENA 
selects
heavier edges 
for $S(t)$ from both $R(t)$ with $S(t-1)$,
} 
so that the weight of $S(t)$ is larger than or equal to those of both $R(t)$ and $S(t-1)$. 
We color-code and orient edges of $R(t)$ and $S(t-1)$, like in~\cite{GiacconePrabhakarShah2003SerenaJSAC},
as follows. We color all edges in $R(t)$ red and all edges in $S(t-1)$ green, and hence in the sequel,
rename $R(t)$ to $S_r$ (``r" for red) and $S(t-1)$ to $S_g$ (``g" for green) to emphasize the coloring. 
{\it We drop the henceforth unnecessary term $t$ here with the implicit understanding that the focus is on 
the MERGE procedure at time slot $t$.} We also orient all edges in $S_r$ as pointing from input ports (\ie $I$)
to output port (\ie $O$) and all edges in $S_g$ as pointing from output ports to input ports. We use 
notations $S_r(I\!\rightarrow\! O)$ and $S_g(O\!\rightarrow\! I)$ to emphasize this orientation when necessary in the sequel.
Finally, we drop the term $t$ from $S(t)$ and denote the final outcome of the MERGE procedure as $S$. 
\deletedHL{
An example pair of thus oriented full matchings $S_r(I\!\rightarrow\! O)$ and $S_g(O\!\rightarrow\! I)$, over an $8 \!\times \!8$ crossbar, 
are shown in~\autoref{fig: red-green-matching}.
}

Now we describe how the two color-coded oriented full matchings $S_r(I\!\rightarrow\!O)$ and $S_g(O\!\rightarrow\!I)$ are merged to produce the 
final full matching $S$.  The MERGE procedure consists of two steps.
The first step is to simply union the two full matchings, viewed as two subgraphs of the complete bipartite graph $G(I\bigcup O)$, into one that we 
call the {\it union graph} and denote as $S_r(I\!\rightarrow\!O) \bigcup S_g(O\!\rightarrow\! I)$ (or $S_r\bigcup S_g$ in short).
In other words, the union graph $S_r(I\!\rightarrow\! O) \bigcup S_g(O\!\rightarrow\! I)$ 
contains the directed edges 
in both $S_r(I\!\rightarrow\! O)$ and $S_g(O\!\rightarrow\! I)$. 

It is a mathematical fact that any such union graph can be decomposed
into disjoint directed cycles~\cite{GiacconePrabhakarShah2003SerenaJSAC}.  
Furthermore, each directed cycle, starting from an input port $I_i$ and going back to itself, is an alternating path
between a red edge in $S_r$ and a green edge in $S_g$, and hence contains equal numbers of red edges and green edges. 
In other words, this cycle consists of a red sub-matching of $S_r$ and a green sub-matching of $S_g$. 
Then in the second step, for each directed cycle, the MERGE procedure compares the weight of the red sub-matching (\ie the total weight of the red edges
in the cycle), with that of the green sub-matching, and includes the heavier sub-matching in the final merged matching $S$.

\begin{figure}
\begin{minipage}{0.5\textwidth}
\centering 
\includegraphics[width=0.9\textwidth]{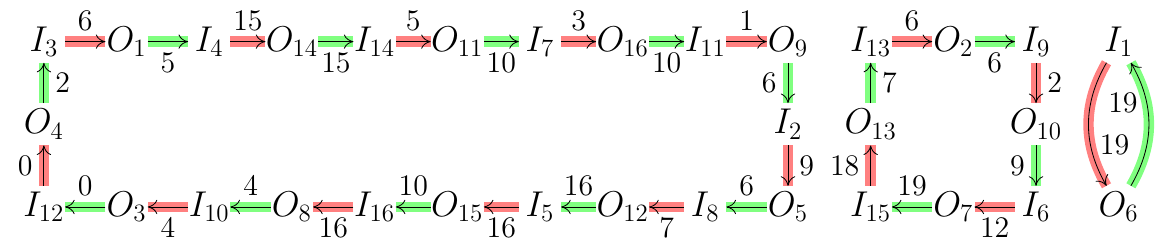}
\caption{Cycles in $S_r(I \!\rightarrow\! O)\bigcup S_g(O \!\rightarrow\! I)$: Edges with red 
 and green shadows are from $S_r(I \!\rightarrow\! O)$ and $S_g(O \!\rightarrow\! I)$ respectively.}\label{fig: cycles-in-union}
\end{minipage}
\\
\begin{minipage}{0.5\textwidth}
\centering 
\includegraphics[width=0.9\textwidth]{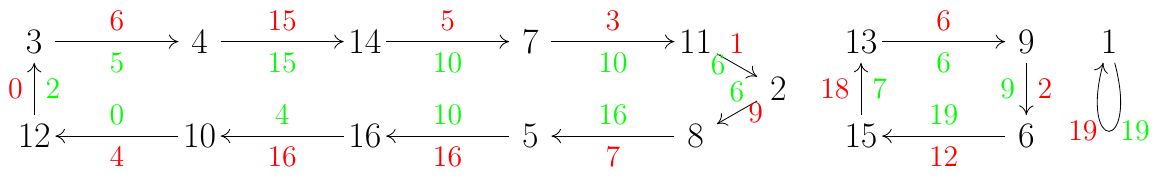}
\caption{Combinatorial Cycles correspond to the cycles in~\autoref{fig: cycles-in-union}.}\label{fig: comb-cycles}
\end{minipage}
\end{figure}

\deletedHL{\autoref{fig: cycles-in-union} shows the union graph of the two full matchings shown in~\autoref{fig: red-green-matching}.}
\medskip
\noindent
{\bf Illustrative Example.}
To illustrate the MERGE procedure by an example, 
\autoref{fig: cycles-in-union} shows the union graph of the following two full matchings over 
a $16\times 16$ bipartite graph (crossbar). 
\modifiedHL{
The number around each edge is its weight. 
}
The first full matching $S_r(I\!\rightarrow\!O)$, written as a permutation with input port numbers ($1$ as $I_1$, $2$ as $I_2$, and so on) at the top
and output port numbers at the bottom ($1$ as $O_1$, $2$ as $O_2$, and so on), is
$\begin{pmatrix}1 & 2 & 3 & 4 & 5 & 6 & 7 & 8 & 9 & 10 & 11 & 12 & 13 & 14 & 15 & 16 \\ 6 & 5 & 1 & 14 & 15 & 7 & 16 & 12 & 10 & 3 & 9 &4 & 2 & 11 & 13 & 8\end{pmatrix}$.
We denote this permutation as $\sigma_r$.  For example, $\sigma_r$ mapping (input port) $3$ to (output port ) $1$ corresponds to the red edge $I_3\!\rightarrow\!O_1$ in \autoref{fig: cycles-in-union} ({\it i.e.},
$I_3$ pairing with $O_1$ in $S_r$, the ``red'' matching).   
The second full matching $S_g(O\!\rightarrow\! I)$, written as a permutation with output port numbers at the top
and input port numbers at the bottom, is
$\begin{pmatrix}1 & 2 & 3 & 4 & 5 & 6 & 7 & 8 & 9 & 10 & 11 & 12 & 13 & 14 & 15 & 16 \\4 & 9 & 12 & 3 & 8 & 1 & 15 & 10 & 2 & 6 & 7 & 5 & 13 & 14 & 16 & 11 \end{pmatrix}$.
We denote this permutation as $\sigma^{-1}_g$.
The union graph contains
three disjoint directed cycles that are of lengths $22, 8, 2$ 
respectively. 
\modifiedHL{
Now, we illustrate the MERGE procedure on the leftmost cycle in~\autoref{fig: cycles-in-union}. 
It is not hard to check that the total weight of the red sub-matching in this cycle 
is $82$ and that of the green sub-matching is $84$. 
Then, the heavier sub-matching, {\it i.e.,} the green one, 
is included into the final merged matching $S$. 
}

The standard centralized algorithm for implementing the MERGE procedure is to linearly traverse every cycle once, by following the directed edges in the cycle,
to obtain the weights of the green and the red sub-matchings that comprise the cycle~\cite{GiacconePrabhakarShah2003SerenaJSAC}. 
Clearly, this algorithm has a computational complexity of $O(N)$.
The primary contribution of SERENADE is to reduce this complexity to $O(\log N)$ per input/output port through parallelization.




\subsection{A Combinatorial View of MERGE}
\label{subsec: merge-perm}

\modifiedHL{
In this section, to better describe MERGE under SERENADE however, we introduce a combinatorial view of MERGE,
}
through which 
the MERGE procedure can be very succinctly characterized
by a single permutation $\sigma \triangleq \sigma_g^{-1} \circ
\sigma_r$, the composition of the two aforementioned full matchings $S_r(I\rightarrow O)$ and $S_g(O\rightarrow I)$ written as permutations.
We do so using the example shown in
~\autoref{fig: cycles-in-union}.
It is not hard to verify that, in this example, 
$\sigma \triangleq \sigma_g^{-1} \circ \sigma_r \!=\! 
 \begin{pmatrix}1 & 2 & 3 & 4 & 5 & 6 & 7 & 8 & 9 & 10 & 11 & 12 & 13 & 14 & 15 & 16\\ 1 & 8 & 4 & 14 & 16 & 15 & 11 & 5 & 6 & 12 & 2 & 3 & 9 & 7 & 13 & 10\end{pmatrix}$. 
We then decompose this permutation $\sigma$ into disjoint combinatorial cycles.
In this example, $\sigma \!=\! (3\,4\,14\,7\,11\,2\,8\,5\,16\,10\,12)(13\,9\,6\,15)(1)$, 
and its {\it cycle decomposition graph},
which contains precisely these three combinatorial cycles, is shown in~\autoref{fig: comb-cycles}.
Note there is a one-to-one correspondence between the graph cycles (of the union graph $S_r\bigcup S_g$) shown in~\autoref{fig: cycles-in-union}
and the combinatorial cycles 
of $\sigma$ shown in~\autoref{fig: comb-cycles}. 
For example, 
cycle 
$\walk{I_{3}, O_1, I_4, O_{14}, \cdots, O_3, I_{12}, O_4, I_3}$,
\modifiedHL{
the leftmost cycle in~\autoref{fig: cycles-in-union}, 
corresponds to the 
leftmost combinatorial cycle $(3\,4\,14\,7\,11\,2\,8\,5\,16\,10\,12)$ in~\autoref{fig: comb-cycles}. 
}
Note that two consecutive edges -- one belonging to the red matching $S_r$ and the other to the green matching $S_g$ -- on the graph cycle  
``collapse" into an edge on the corresponding combinatorial cycle.  
For example, two directed edges 
$(I_{3}, O_1)$ ($\in S_r$) and $(O_1, I_4)$ ($\in S_g$) 
in~\autoref{fig: cycles-in-union} 
collapsed into the directed edge from (input port) 
$3$ to (input port) $4$ in the combinatorial cycle $(3\,4\,14\,7\,11\,2\,8\,5\,16\,10\,12)$  
in~\autoref{fig: comb-cycles}. 
Hence each combinatorial cycle subsumes
a red sub-matching and a green sub-matching that collapse into it. 
Note also that each vertex on the cycle decomposition graph corresponds to an input port. For example, vertex 
``3'' in $(3\,4\,14\,7\,11\,2\,8\,5\,16\,10\,12)$ in~\autoref{fig: comb-cycles} 
corresponds to 
input port 
$I_{3}$ in~\autoref{fig: cycles-in-union}. 
Hence, we use the terms ``vertex'' and ``input port'' interchangeably in the sequel. 

We assign a green weight $w_g(\cdot)$ and 
a red weight $w_r(\cdot)$ -- to each combinatorial edge $e$ in the cycle decomposition graph -- that are equal to the respective weights of the green and the red edges that 
collapse into $e$. 
\modifiedHL{
For example, the green and red numbers around each edge shown in~\autoref{fig: comb-cycles} represent 
its green and red weights respectively. 
}
We also define the green (or red) weight of a combinatorial cycle as the total green 
(or red) weight of all combinatorial edges on the cycle.
Clearly, this green (or red) weight is equal to the weight of the green (or red) sub-matching this cycle subsumes. 
\modifiedHL{
Under this combinatorial view, the MERGE procedure of SERENA can be stated 
as follows:
}
{\it For each combinatorial cycle \deletedHL{in the cycle decomposition graph}of 
$\sigma$, we compare its red weight with its green weight, and include in $S$ the corresponding heavier sub-matching.}

\subsection{Walks on Cycles}
\label{subsec: walks-on-cycles}



\modifiedHL{
Finally, we introduce the concept of walk on a cycle decomposition
graph. It greatly simplifies the descriptions of SERENADE, as it will become 
clear later that SERENADE is all 
about how to emulate SERENA using information, each input port obtains, regrading a few walks with  
lengths of power of $2$.  
} 
Recall that a {\it walk} in a general graph $G(V,E)$ is an ordered sequence of vertices, $\walk{v_1,v_2,\cdots,v_k}$ such that $(v_j,v_{j+1}) \!\in\! E$ for any $j \!\in\! \{1,2,\dots,k-1\}$;  note that a {\it walk}, unlike a 
{\it path}, can traverse a vertex or edge more than once.  Clearly, in the cycle decomposition graph of $\sigma$, every walk (say starting from a vertex $i$) 
circles around a combinatorial cycle (the one that $i$ lies on), and hence necessarily takes the following form: 
$\walk{i , \sigma(i) , \sigma^2(i) , \cdots , \sigma^m(i)}$.
For notational convenience, we denote this walk as $i \leadsto \sigma^{m}(i)$.
For example, with respect to 
the combinatorial cycle 
$(3\,4\,14\,7\,11\,2\,8\,5\,16\,10\,12)$
in~\autoref{fig: comb-cycles}, the walk $3\leadsto \sigma^8(3)$
represents 
$\walk{3, 4, 14, 7, 11, 2, 8, 5, 16}$,  
which consists of 
$8$ directed edges on the cycle. 

Generalizing this notation (of a walk), we define $\sigma^{m_1}(i) \!\leadsto\! \sigma^{m_2}(i)$ as the $(m_2\!-\!m_1)$-edge-long walk $\walk{\sigma^{m_1}(i),\sigma^{(m_1\!+\!1)}(i), \cdots, \sigma^{m_2}(i)}$, 
where $m_1 \!<\! m_2$ are integers, and both $m_1$ and $m_2$ could be negative.
We define its red weight, denoted as $w_r(\sigma^{m_1}(i) \!\leadsto\! \sigma^{m_2}(i))$, as the sum of the red weights of all edges in $\sigma^{m_1}(i) \!\leadsto\! \sigma^{m_2}(i)$.  Note that if an edge is traversed multiple times in a walk, the red weight of the edge is accounted for multiple times.  The green weight
of the walk, denoted as $w_g(\sigma^{m_1}(i) \!\leadsto\! \sigma^{m_2}(i))$, is similarly defined. 
\section{Overview of SERENADE}
\label{sec: serenade}


\modifiedHL{
In this section, we provide a high-level overview of SERENADE.  We first introduce the core idea of SERENADE that is based on an important concept called ``discover". 
Then, we give a high-level description of two algorithmic stages of SERENADE: knowledge-discovery 
stage and distributed binary search stage. 
}\movedHL{
For ease of presentation (\textit{e.g.,} no need to put floors or ceilings around each 
occurrence of $\log_2 N$), we have assumed that $N$
is a power of $2$ throughout this paper;  
}\modifiedHL{
SERENADE works just as well when $N$ is not. 
}

\subsection{Core Idea of SERENADE}\label{subsec: design-guideline}

\modifiedHL{
The core idea of SERENADE is for all vertices on a combinatorial cycle, or a designated vertex among them, to {\it discover} (defined next)
itself or another vertex on the same cycle twice.  As will be shown next in \autoref{fact:disc}, when this happens, these vertices will know precisely whether the green weight or the red weight 
of the cycle is larger, and hence will select the same heavier sub-matching as they would under SERENA.  If this happens on every combinatorial cycle, then SERENADE exactly emulates SERENA.
}


\modifiedHL{
\begin{definition}\label{def:disc}
Given two vertices $i, j$ in any combinatorial cycle of $\sigma$, 
we say that vertex $i$ {\it discovers} vertex $j$ if $i$ learns the identity of $j$ and 
the weights of a walk from $i$ to $j$ or from $j$ to $i$. 
\end{definition}

By this definition, every vertex $i$ discovers itself, once at the beginning ({\it i.e.}, before any algorithmic steps), via the empty (0-edge-long) walk from $i$ to $i$.


\begin{lemma}[Property of ``Discover'']\label{fact:disc}\label{lemma:disc}
Let $i$ and $j$ be two vertices, which may or may not be the same vertex, on a combinatorial cycle of $\sigma$.  If $i$ discovers $j$ 
twice via two different walks, then vertex $i$ knows precisely whether the green weight or the red weight 
of the cycle is larger. 
\end{lemma}

\begin{IEEEproof}
See~\autoref{app:proof-lemma-01}.
\end{IEEEproof}
}

\subsection{High-Level Description of SERENADE}\label{subsec:high-level-serenade}

\modifiedHL{
As mentioned earlier, SERENADE consists of two algorithmic stages:  knowledge-discovery stage and 
distributed binary search stage. In this section, we give the high-level descriptions of 
these two stages, deferring their details to~\autoref{subsec:knowledge-discovery} 
and~\autoref{subsec:bs} respectively.

\medskip
\noindent
{\bf Knowledge-Discovery Stage}.
The knowledge-discovery stage uses the standard technique of two-directional exploration with
successively doubled distance in distributed computing~\cite{Lynch1996Distributed}.
The basic idea of the algorithm is for each vertex $i$ to exchange information, 
during the $k^{th}$ ($1\!\le\!k\!\le\!\log_2 N$) iteration\footnote{As will be shown in~\autoref{subsec:knowledge-discover-procedure}, there is a $0^{th}$ iteration at the beginning, with which each vertex $i$ discovers $\sigma(i)$ and $\sigma^{-1}(i)$.},
with vertices ``$(\pm 2^{k-1})$ $\sigma$-hops" away (\ie $\sigma^{2^{k-1}}(i)$ 
and $\sigma^{-(2^{k-1})}(i)$) 
to discover two vertices ``$(\pm 2^k)$ $\sigma$-hops" away (\ie $\sigma^{2^k}(i)$ 
and $\sigma^{-(2^k)}(i)$).  If either of the two 
vertices has been discovered twice by $i$, then, by~\autoref{fact:disc}, we know that 
vertex $i$ can make the same matching 
decision as it would under SERENA. 
In the example shown in~\autoref{fig: comb-cycles}, vertex $3$ communicates, during the $1^{st}$ iteration, with 
vertices $4\!=\!\sigma(3)$ and $12\!=\!\sigma^{-1}(3)$ to discover vertices
$14\!=\!\sigma^{2}(3)$ and $10\!=\!\sigma^{-2}(3)$, and communicates, during the $2^{nd}$ 
iteration, with the newly-discovered vertices $14$ and $10$ to discover vertices $11=\sigma^{4}(3)$ 
and $5\!=\!\sigma^{-4}(3)$, and so on in the 
next $(\log_2 N)\!-\! 2$ iterations.  


\medskip
\noindent
{\bf Distributed Binary Search Stage}. 
After the $1\!+\!\log_2 N$ iterations of the knowledge discovery, a vertex $i$, residing on a cycle, 
will discover a vertex on the same cycle 
twice and hence make the same matching decision as it would under SERENA, 
if the cycle is ouroboros (to be defined in~\autoref{subsec:ouroborous-cycle}).
However, not all cycles are ouroboros, as will be shown in~\autoref{subsec:ouroborous-cycle}.   
Those and only those vertices, residing on non-ouroboros cycles, then perform an additional distributed binary 
search, the purpose of which is to let 
a designated vertex in each non-ouroboros cycle discover itself for a second time.
We will show 
in~\autoref{sec:leader-election} that the elections of those designated vertices ({\it i.e.,} leader election) can be 
seamlessly embedded into the $1\!+\!\log_2N$ iterations of the knowledge-discovery procedure. 
We will show in~\autoref{sec:bs} that the distributed binary search finishes 
in at most $\log_2 N$ iterations. After the distributed binary search, each designated vertex informs 
the switch controller whether the green or the red sub-matching should be selected on the non-ouroboros cycle it resides on. 
The switch controller 
then broadcasts these decisions to all $N$ vertices, and every vertex on a non-ouroboros cycle 
will carry out the corresponding matching decision.

\autoref{thm:serenade-exact-emu} is a main result of this paper.  Its proof is straightforward after we have
proved the correctness of the knowledge-discovery procedure (\autoref{subsec:knowledge-discover-procedure}) and the distributed binary 
search (\autoref{sec:bs}).

\begin{theorem}\label{thm:serenade-exact-emu}
SERENADE exactly emulates SERENA~\cite{GiacconePrabhakarShah2003SerenaJSAC} within 
$O(\log N)$ iterations. More precisely, at most $1\!+\!\log_2N$ iterations are needed for the knowledge discovery 
procedure and at most $\log_2 N$ iterations for the distributed binary search.
\end{theorem}
} 










\section{Knowledge-Discovery Stage}\label{subsec:knowledge-discovery}\label{sec:knowledge-discovery}

\modifiedHL{
In this section, we describe the details of the knowledge-discovery stage. We start with describing the 
information obtained by the knowledge-discovery procedure in~\autoref{subsubsec:knowledge-sets}; 
the detailed algorithmic steps in each iteration will be described later in~\autoref{subsec:knowledge-discover-procedure}. 
In~\autoref{subsec: complexity-common}, we analyze the time and message complexities of the knowledge-discovery 
procedure. Finally, we explain in~\autoref{subsec:ouroborous-cycle} which vertices can discover some vertex twice during the knowledge-discovery procedure 
by introducing the concept of ``ouroboros cycle''.
}

\subsection{Knowledge Sets}\label{subsubsec:knowledge-sets}\label{subsec:knowledge-sets}

\modifiedHL{
We will show next that, for $0\!\le\!k \!\le\!\log_2 N$ (there is a $0^{th}$ iteration at the beginning), after the $k^{th}$ iteration, each vertex $i$ learns the following two 
knowledge sets:  $\phi^{(i)}_{k+}$ and $\phi^{(i)}_{k-}$.
}\movedHL{
Knowledge set $\phi^{(i)}_{k+}$ contains three quantities concerning the 
vertex (input port)
that is $2^k$ $\sigma$-hops ``downstream"
({\it w.r.t.} the ``direction'' of $\sigma$), relative to vertex $i$, in the cycle decomposition graph of $\sigma$:
\begin{spenumerate}
    \item $\sigma^{2^k}(i)$, the identity of that vertex,
    \item $w_r(i\leadsto\sigma^{2^k}(i))$, the red weight of the $2^k$-edge-long walk from $i$ to that 
    vertex, and
    \item $w_g(i\leadsto\sigma^{2^k}(i))$, the green weight of the walk.
\end{spenumerate}
}

\movedHL{
Similarly, knowledge set $\phi^{(i)}_{k-}$ contains the three quantities concerning the 
vertex that is $2^k$ $\sigma$-hops ``upstream" relative to vertex $i$, namely
$\sigma^{-2^k}(i)$, $w_r(\sigma^{-2^k}(i)\leadsto i)$, and $w_g(\sigma^{-2^k}(i)\leadsto i)$.  
For example, after the $3^{rd}$ iteration, vertex $3$ learns the identities of $16\!=\!\sigma^8(3)$ and $7\!=\!\sigma^{-8}(3)$ (vertices ``$(\pm 8)$ $\sigma$-hops" away) and the green and the red weights of the walks $3\!\leadsto\!\sigma^8(3)$ and $\sigma^{-8}(3)\!\leadsto \!3$.
}

\modifiedHL{
As we will show in~\autoref{subsec:knowledge-discover-procedure}, the knowledge-discovery procedure might halt 
before finishing the $1 \!+\!\log_2N$ iterations, so each vertex learns at most $2\!+\!2\log_2N$ 
knowledge sets during the knowledge-discovery procedure. 
Note the $2\!+\!2\log_2N$ knowledge sets 
are a tiny percentage of information
the switch controller has under SERENA:  
}\existingHL{
the former scales as $O(\log N)$ whereas the latter scales as $O(N)$.  
For example, in~\autoref{fig: comb-cycles}, vertex $3$ knows only the values the permutation
function $\sigma(\bigdot)$ takes on argument values $3$, $4$, and $12$, whereas under SERENA the (central) switch controller would know that on all $N \!=\! 16$ argument values.  
In general, 
different vertices have very different sets of such partial knowledge under SERENADE.  
For example, vertex $2$ knows only the values the permutation
function $\sigma(\bigdot)$ takes on argument values $2$, $8$, and $11$.     
However, despite this ``blind men (different vertices) and an elephant ($\sigma$ and the green and the red weights of all walks
on the combinatorial cycles of $\sigma$)'' situation, these vertices manage to collaboratively perform the approximate or the exact  
MERGE operation ({\it i.e.}, making do with less).
}


\subsection{Knowledge-Discovery Procedure}\label{subsec:knowledge-discover-procedure}






\modifiedHL{
We now describe the $1 \!+\! \log_2 N$ iterations of the knowledge-discovery procedure in detail and explain how
these iterations allow every vertex $i$ to concurrently obtain its knowledge sets 
$\phi^{(i)}_{k+}$ and $\phi^{(i)}_{k-}$ for $0\!\le\! k \!\le\! \log_2 N$. The pseudocode of the knowledge-discovery 
procedure at vertex $i$ is presented in~\autoref{alg: serenade-general}, that executed at any other vertex is identical.
}

\begin{algorithm}[t]

\SetAlFnt{\small}
\caption{\small Knowledge-discovery procedure at vertex $i$.}
\label{alg: serenade-general}
\SetKwData{Left}{left}\SetKwData{This}{this}\SetKwData{Up}{up}
\SetKwFunction{Union}{Union}\SetKwFunction{FindCompress}{FindCompress}
\SetKwInOut{Input}{global}\SetKwInOut{Output}{output}



\For{$k\leftarrow 0$ \KwTo $\log_2 N$}{
\eIf{$k=0$}{
\tcp{\small The $0^{th}$ iteration}

Receive from output port $o_r$ pairing with $i$ in $S_r$: identity $\sigma(i)$ and weight \scalebox{0.9}{$w_g(i\!\rightarrow\! \sigma(i))$}\;\label{kd:receive-0}

Receive from output port $o_g$ pairing with $i$ in $S_g$: identity $\sigma^{\!-\!1}(i)$ and weight \scalebox{0.9}{$w_r(\sigma^{\!-\!1}(i)\!\rightarrow\! i)$}\;\label{kd:receive-0-minus}

Compute knowledge set $\phi_{k+}^{(i)}$: \scalebox{0.9}{$\phi_{k+}^{(i)}\!\leftarrow\! \{\sigma(i), w_r(i\!\rightarrow\! \sigma(i)), w_g(i\!\rightarrow\! \sigma(i))\}$}\;\label{kd:phi-0}

Compute knowledge set $\phi_{k-}^{(i)}$ similarly\;\label{kd:phi-0-minus}


}{
\tcp{\small The subsequent iterations}
$i_D\leftarrow \sigma^{2^{k-1}}(i)$\;
$i_U\leftarrow \sigma^{-2^{k-1}}(i)$\;

Send to $i_U$ the knowledge set $\phi^{(i)}_{(k-1)+}$\;
\label{serenade: send-up}

Receive from $i_D$ the knowledge set $\phi^{(i_D)}_{(k-1)+}$\;
\label{serenade: receive-down}

Send to $i_D$ the knowledge set $\phi^{(i)}_{(k-1)-}$\;
\label{serenade: send-down}

Receive from $i_U$ the knowledge set $\phi^{(i_U)}_{(k-1)-}$\;
\label{serenade: receive-up}

Compute knowledge set $\phi^{(i)}_{k}$ with $\phi^{(i)}_{(k-1)+}$ and $\phi^{(i_D)}_{(k-1)+}$ by~\autoref{eq:comp-know}\;
\label{serenade: compute-knowledge-set}

Compute knowledge set $\phi^{(i)}_{k-}$ similarly with $\phi^{(i)}_{(k-1)-}$ and $\phi^{(i_U)}_{(k-1)-}$\;
\label{serenade: compute-knowledge-set-minus}
}
\tcp{Halt checking}
{\bf Halt} if vertex $i$ discovers any vertex twice (in light of newly discovered vertices $\sigma^{2^k}(i)$ and 
$\sigma^{-2^k}(i)$)\;\label{serenade: halt-check}


}

\end{algorithm}

\subsubsection{The \texorpdfstring{$0^{th}$}{0th} Iteration}\label{subsubsec: iteration-0}
\movedHL{
We start with describing the $0^{th}$ iteration, the operation of which is slightly different than that of subsequent iterations
in that whereas messages are exchanged only between input ports in all subsequent iterations, 
}\modifiedHL{
messages are also exchanged between input ports and output ports in the $0^{th}$ iteration.  
Suppose input port $i$ is paired with output port $o_r$ in the (red) full matching $S_r$ and with
output port $o_g$ in the (green) full matching $S_g$.  
The $0^{th}$ iteration contains two rounds of message exchanges. 
In the first round, input port $i$ receives from output port $o_r$
a message which includes the identity $\sigma(i)$ and the weight of edge $o_r\!\rightarrow \sigma(i)\!$, {\it i.e.,} the green weight of edge 
$i \!\rightarrow\! \sigma(i)$ (\autoref{kd:receive-0}). Note that input port $\sigma(i)$ is paired with $o_r$ in $S_g$, so 
$o_r$ knows the identity of $\sigma(i)$ and the weight of edge $o_r\!\rightarrow \sigma(i)\!$. 
With the newly received information, input port $i$ can calculate knowledge set $\phi_{0+}^{(i)}$ (\autoref{kd:phi-0}). 
For example, in this round, input port $3$ in~\autoref{fig: cycles-in-union} receives from output port $1$ the identity of 
input port $4$ and the weight, which is $5$, of edge $O_1\!\rightarrow\!I_4$. 
Similarly, in the second round, 
input port $i$ receives from output port $o_g$
a message which includes the identity $\sigma^{-1}(i)$ and the weight of edge $\sigma^{-1}(i)\!\rightarrow \!o_g$, {\it i.e.,} the red weight of edge 
$\sigma^{-1}(i) \!\rightarrow\! i$ (\autoref{kd:receive-0-minus}). Note that input port $\sigma^{-1}(i)$ is paired with $o_g$ in $S_r$. 
With the newly received information, input port $i$ can calculate knowledge set $\phi_{0-}^{(i)}$ (\autoref{kd:phi-0-minus}). For example,  
in this round, input port $3$ in~\autoref{fig: cycles-in-union} receives from output port $O_4$ the identity of input port $12$ and the weight, 
which is $0$, of edge $I_{12}\!\rightarrow\!O_4$. 
Therefore, after this $0^{th}$ iteration, each input port $i$ 
obtains the knowledge sets $\phi^{(i)}_{0+}$ and $\phi^{(i)}_{0-}$, or
in other words, discovers $\sigma(i)$ and $\sigma^{-1}(i)$. }

\subsubsection{Subsequent Iterations}\label{subsubsec: subsequent-iterations}

\modifiedHL{
The subsequent iterations
can be described inductively as follows. 
}\modifiedHL{
Suppose after iteration $k\!-\!1$ (for any $1\!\le\! k\!\le\! \log_2 N$), every vertex $i$ discovers 
its upstream vertex $i_U \!=\! \sigma^{-2^{k-1}}(i)$ and its downstream vertex $i_D \!=\! \sigma^{2^{k-1}}(i)$.
}\modifiedHL{
Lines \ref{serenade: send-up}-\ref{serenade: compute-knowledge-set-minus} in \autoref{alg: serenade-general} show 
how vertex $i$ discovers 
$\sigma^{2^{k}}(i)$ and $\sigma^{-2^{k}}(i)$ via two rounds of message exchanges of the 
iteration $k$. 
}\modifiedHL{
In the first round, 
vertex $i$ sends the 
knowledge set $\phi^{(i)}_{(k-1)+}$, obtained during iteration $k\!-\!1$,
to the upstream vertex $i_U$ (\autoref{serenade: send-up}).
Meanwhile, vertex $i$ receives from the downstream vertex $i_D$ 
its knowledge set $\phi^{(i_D)}_{(k-1)+}$ (\autoref{serenade: receive-down}), which, 
as explained earlier, contains the values of 
$\sigma^{2^{k-1}}(i_D)$, $w_r(i_D \!\leadsto\! \sigma^{2^{k-1}}(i_D))$, 
and $w_g(i_D \!\leadsto\! \sigma^{2^{k-1}}(i_D))$. Having obtained these three values, vertex $i$ pieces together its knowledge set
$\phi^{(i)}_{k+}$ (\autoref{serenade: compute-knowledge-set}) as follows.
}\movedHL{
\setlength{\belowdisplayskip}{0pt} \setlength{\belowdisplayshortskip}{0pt}
\setlength{\abovedisplayskip}{4pt} \setlength{\abovedisplayshortskip}{4pt}
\begin{equation}\label{eq:comp-know}\resizebox{0.913\hsize}{!}{
$\begin{cases}
\sigma^{2^k}(i) &\leftarrow \sigma^{2^{k-1}}\bigParenthes{i_D}\\
w_r\bigParenthes{i \leadsto \sigma^{2^k}(i)} &\leftarrow
w_r\bigParenthes{i \leadsto \sigma^{2^{k-1}}(i)} + w_r\bigParenthes{i_D \leadsto \sigma^{2^{k-1}}(i_D)}\\
w_g\bigParenthes{i \leadsto \sigma^{2^k}(i)} &\leftarrow
w_g\bigParenthes{i \leadsto \sigma^{2^{k-1}}(i)} + w_g\bigParenthes{i_D \leadsto \sigma^{2^{k-1}}(i_D)}
\end{cases}$
}\end{equation}
}\modifiedHL{
Note that vertex $i$ already knows $\phi^{(i)}_{(k-1)+}$, which includes
$w_r(i \leadsto \sigma^{2^{k-1}}(i))$ and $w_g(i \leadsto \sigma^{2^{k-1}}(i))$.
}

\movedHL{
Similarly, in the second round of message exchanges, vertex $i$ sends 
$\phi^{(i)}_{(k-1)-}$ to the downstream vertex $i_D$ (\autoref{serenade: send-down}), 
}\modifiedHL{
and meanwhile receives
$\phi_{(k-1)-}^{(i_U)}$ from the upstream vertex $i_U$ (\autoref{serenade: receive-up}). 
}
The latter 
knowledge set (\ie $\phi_{(k-1)-}^{(i_U)}$), combined with the 
knowledge set
$\phi_{(k-1)-}^{(i)}$ that vertex $i$ already knows, 
allows $i$ to piece together the 
knowledge set $\phi_{k-}^{(i)}$. 
\modifiedHL{
Therefore, vertex $i$ obtains $\phi_{k+}^{(i)}$ and $\phi_{k-}^{(i)}$, or in other words discovers  
$\sigma^{2^k}(i)$ and $\sigma^{-2^k}(i)$, after the $k^{th}$ iteration. 
}

\begin{figure}
\centering
\includegraphics[width=0.45\textwidth]{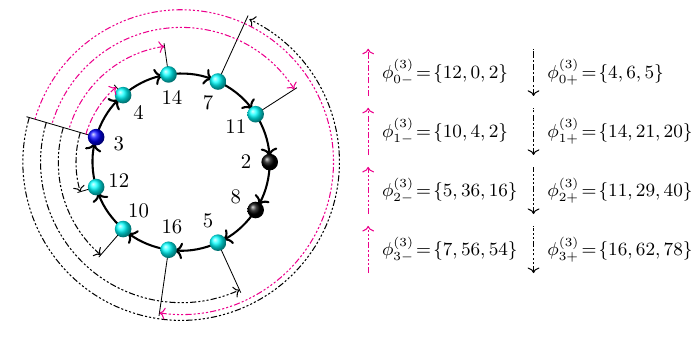}
\caption{Illustration of the knowledge-discovery procedure: messages sent by vertex $3$ in~\autoref{fig: comb-cycles}.}
\label{fig:knowledge-disc}
\end{figure}

\modifiedHL{
\medskip
\noindent
{\bf Illustrative Example}. \autoref{fig:knowledge-disc} presents the messages sent by vertex $3$ 
in~\autoref{fig: comb-cycles} during the $k^{th}$ ($1\!\le\!k\!\le\!4$) iteration of the 
knowledge-discovery procedure. 
For example, in the $3^{rd}$ iteration, vertex $3$ sends to vertex $11\!=\!\sigma^4(3)$ 
the knowledge set $\phi_{2-}^{(3)}\!=\!\{\sigma^{-4}(3),w_r(\sigma^{-4}(3)\!\leadsto\!3), w_g(\sigma^{-4}(3)\!\leadsto\!3)\}\!=\!\{5,36, 16\}$ (the $3^{rd}$ up arrow from top 
to bottom in right half of~\autoref{fig:knowledge-disc}) 
it learns during the $2^{nd}$ iteration. 
It also sends to vertex $5 \!=\! \sigma^{-4}(3)$ the knowledge set 
$\phi_{2+}^{(3)} \!=\!\{\sigma^{4}(3),w_r(3\!\leadsto\!\sigma^{4}(3)),w_g(3\!\leadsto\!\sigma^{4}(3))\}\!=\!\{11,29,40\}$ (the $3^{rd}$ 
down arrow from top to bottom in right half of~\autoref{fig:knowledge-disc}). 
Though it is not shown in~\autoref{fig:knowledge-disc}, 
in the same iteration, vertex $3$ 
receives $\phi_{2+}^{(11)}=\{\sigma^4(11),w_r(11\!\leadsto\!\sigma^4(11)), w_g(11\!\leadsto\!\sigma^4(11))\}\!=\!\{16,33,38\}$  from vertex $11\!=\!\sigma^4(3)$ 
so that it can compute, by~\autoref{eq:comp-know},  
$\sigma^8(3)\!=\!16$, $w_r(3\!\leadsto\!\sigma^8(3))\!=\!w_r(3\!\leadsto\!\sigma^{4}(3))\!+\!w_r(11\!\leadsto\!\sigma^4(11))\!=\!29+33\!=\!62$, and $w_g(3\!\leadsto\!\sigma^8(3))\!=\!w_g(3\!\leadsto\!\sigma^{4}(3))\!+\!w_g(11\!\leadsto\!\sigma^4(11))\!=\!40\!+\!38\!=\!78$, which is precisely $\phi_{3+}^{(3)}$. 
It also receives $\phi_{2-}^{(5)} \!=\!\{\sigma^{-4}(5),w_r(\sigma^{-4}(5)\!\leadsto\!5),w_g(\sigma^{-4}(5)\!\leadsto\!5)\}$ from vertex $5 = \sigma^{-4}(3)$. 
Similarly, it can compute $\phi_{3-}^{(3)}\!=\!\{\sigma^{-8}(3),w_r(\sigma^{-8}(3)\!\leadsto\!3),w_g(\sigma^{-8}(3)\!\leadsto\!3)\}\!=\!\{7,56,54\}$. 
Therefore, vertex $3$ discovers vertex $16\!=\!\sigma^8(3)$ and $7\!=\!\sigma^{-8}(3)$ respectively. 
}

\modifiedHL{
\subsubsection{Early Halt Checking}
The knowledge-discovery procedure might halt before finishing the $1\!+\!\log_2 N$ iterations. 
As shown in~\autoref{serenade: halt-check}, the procedure will halt 
if vertex $i$ discovers some vertex twice. More precisely, if vertex $i$ discovers the same 
vertex twice in the same iteration, {\it i.e.,} $\sigma^{2^k}(i)\!=\!\sigma^{-2^k}(i)$ 
or it discovers a vertex that has been discovered in the previous iterations, {\it i.e.,} 
vertex $i$ has already discovered $\sigma^{2^k}(i)$ (or $\sigma^{-2^k}(i)$) in the 
previous iterations. By~\autoref{fact:disc}, 
we can conclude that vertex $i$ can make the exact same matching decision as it would under SERENA. 

\medskip
\noindent
{\bf Halt Checking in $O(1)$ Per Iteration.} 
Vertex $i$ can finish halt checking in $O(1)$ (per port) per iteration, 
or $O(\log N)$ in total, using a pointer array $B[1 .. N]$. Here, we only need to 
show that the latter case described above, {\it i.e.,} checking whether vertex $i$ has already 
discovered $\sigma^{2^k}(i)$ (or $\sigma^{-2^k}(i)$) in the 
previous iterations, in $O(1)$, as the checking for the 
former case ({\it i.e.,} whether $\sigma^{2^k}(i)\!=\!\sigma^{-2^k}(i)$) is obviously $O(1)$. 
Each array entry $B[i']$ initially points to \code{NULL}. At the end of each 
iteration (including the $0^{th}$ iteration), 
vertex $i$ simply checks whether $B[\sigma^{2^k}(i)]\!\neq\!\code{NULL}$ 
(or $B[\sigma^{-2^k}(i)]\!\neq\!\code{NULL}$).  
If so, then $\sigma^{2^k}(i)$ (or $\sigma^{-2^k}(i)$) has been discovered in the previous iterations. 
Otherwise, we update $B$ as follows: pointing $B[\sigma^{2^k}(i)]$ and $B[\sigma^{-2^k}(i)]$ 
to the knowledge set 
$\phi_{k+}^{(i)}$ and $\phi_{k-}^{(i)}$ respectively. 

Note that we need to reset the values of all $N$ entries of $B$ to $\code{NULL}$ at 
the end of a matching computation.  
The computational complexity of the reset is $O(\log N)$ (instead of $O(N)$) because each 
non-null entry of $B$ is indexed by the identity field of a knowledge set, the total number of 
which is upper-bounded by $2 \!+\! 2\log_2 N $.  
Hence the total computational complexity for each vertex $i$ to 
finish halt checking, {\it i.e.,} \autoref{serenade: halt-check} of~\autoref{alg: serenade-general}, 
is $O(1)$ (per port) per iteration, or $O(\log N)$ in total.
}

\modifiedHL{
\medskip
\noindent
{\bf All or None Lemma.} 
Using the similar operations as in the proof of~\autoref{fact:disc}, vertex $i$ can use $O(1)$ operations to 
decide which is heaver between the green weight and the red weight of the cycle that vertex $i$ belongs to. 
So can other vertices belonging to the same cycle (as vertex $i$) by using the following lemma.

\begin{lemma}[All or None]\label{lemma:all-or-nothing}
During the execution of the knowledge-discovery procedure in SERENADE, if any vertex $i$ halts before 
finishing the $1\!+\!\log_2 N$ iterations, {\it i.e.,} halting because of discovering some vertex 
twice in~\autoref{serenade: halt-check} of~\autoref{alg: serenade-general}, 
then all other vertices belonging to the same cycle will 
also halt in the same iteration 
\end{lemma}
\begin{IEEEproof}
See~\autoref{app:proof-of-lemma-02}.
\end{IEEEproof}
}

\subsubsection{Discussions}\label{subsubsec: logical_bypass}
\movedHL{
In describing 
the knowledge-discovery procedure,
we assume that input ports
can communicate directly with each other.  This is a 
realistic assumption, because in most 
real-world switch products, each line card $i$ is full-duplex in the sense
the logical input port $i$ and the logical output port $i$ are 
co-located in the same physical line card $i$.  In this case, for example, an input port $i_1$
can communicate with another input port $i_2$ by sending information to output port $i_2$, which then relays it to 
the input port $i_2$ through the ``local bypass", presumably at little or no communication 
costs. 
}\modifiedHL{
However, SERENADE also works for the type of switches that do not have such a ``local bypass," 
by letting an output port to serve as a relay, albeit at twice the communication costs.  
}\movedHL{
More precisely, in the example above, the input port $i_1$ can send the 
information first to the output port $i_1$, which then relays the information 
to the input port $i_2$. 
}

\subsection{Complexity Analysis}
\label{subsec: complexity-common}

\modifiedHL{
We now analyze the time and message complexities of the knowledge-discovery procedure.

\medskip
\noindent
{\bf Time Complexity}.
The time complexity of the knowledge-discovery procedure is (at most) $1 \!+\! \log_2 N$ iterations, 
and that of each iteration is 
several operations for local computation for computing knowledge sets (Lines \ref{serenade: compute-knowledge-set}-\ref{serenade: compute-knowledge-set-minus} of \autoref{alg: serenade-general}) and halt checking   
(\autoref{serenade: halt-check} of \autoref{alg: serenade-general}). Clearly, those operations can be performed 
in $O(1)$. 
}

\medskip
\noindent
{\bf Message Complexity}.
The message complexity of the knowledge-discovery procedure is $O(\log N)$ messages per vertex, since 
every vertex needs to send (and receive) two messages 
during each iteration. 
\modifiedHL{
In every message, it suffices to only include 
$w_r(\cdot)\!-\!w_g(\cdot)$, the difference between the red and the green 
weights of the corresponding walk. Therefore, 
each message ({\it i.e.,} knowledge set) can be encoded in $C\!+\!\log_2 N$ bits, where 
$C$ is the maximum number of bits needed to encode this difference. 
}

\subsection{Early Halt: The Ouroboros Cycles}\label{subsec:ouroborous-cycle}

\modifiedHL{
In this section, we define the concept of an {\it ouroboros cycle}, and prove~\autoref{lemma:ouroboros-lemma}, 
which states that all vertices on an ouroboros cycle 
can halt (\autoref{serenade: halt-check} of~\autoref{alg: serenade-general}) 
and make the exact same matching decisions as they would under SERENA, 
without performing the distributed binary search.  
Ouroboros is the ancient Greek symbol depicting a serpent devouring its own tail. We ``borrow'' 
this concept because what happens in ouroboros cycles is very similar to what is depicted by the symbol ``Ouroboros''.

\begin{definition}[Ouroboros Cycle]\label{fact: ouroboros-numbers}\label{def: ouroboros-cycle}
A cycle is said to be {\it ouroboros} if and only if its length $\ell$ is an {\it ouroboros number} ({\it w.r.t.} $N$), 
defined as a positive divisor of a number that takes one of the following three forms: 
(I) $2^\alpha$, (II) $2^\beta \!-\! 2^\gamma$, and (III) $2^\beta \!+\! 2^\gamma$, where $\alpha$, 
$\beta$ and $\gamma$ are nonnegative integers that satisfy $\alpha\!\leq\!\lceil\!\log_2 N \rceil$ and 
$\gamma\!<\!\beta\!\leq\!\lceil\!\log_2 N \rceil$.
\end{definition}

It is not hard to check that, in~\autoref{fig: comb-cycles}, the leftmost cycle (of length $11$) 
is not ouroboros ({\it i.e.}, non-ouroboros), but the other two are. 

{\bf The following lemma shows a nice property of ouroboros cycles, whose proof can be found 
in 
\autoref{app:proof-of-lemma-03}.
}
}

\begin{lemma}[Ouroboros Lemma]\label{lemma:ouroboros-lemma}
Vertex $i$ will discover twice a vertex on the same cycle (as itself) during the knowledge-discovery stage, if it is on an ouroboros cycle.
\end{lemma}

\noindent
{\bf Remark.} {\bf Readers may wonder if we can do away with the distributed binary search 
simply by running a little more iterations (say $0.5\log_2N$ more iterations), because more 
iterations means that more vertices may discover a vertex twice. 
Unfortunately, as shown in 
\autoref{app:why-not-more}, 
there exists some numbers (cycle lengths) that are ``hardcore non-ouroboros" 
in the sense a vertex $i$ on a cycle of such a length $\ell$ needs to 
run exactly $\lceil \ell/2\rceil$ iterations to discover a vertex twice. }

\section{Leader Election}
\label{subsec: leader-election}\label{sec:leader-election}

\modifiedHL{
We have shown that the $1\!+\!\log_2N$ iterations of the knowledge-discovery procedure alone is not enough for 
SERENADE to emulate SERENA exactly. To do so, SERENADE needs an additional distributed binary search. 
As mentioned in~\autoref{sec: serenade}, the distributed binary search requires every non-ouroboros cycle to 
elect a designated vertex, which is decided through a leader election by vertices 
on this cycle. In this section, we describe 
how to embed this leader election seamlessly into the knowledge-discovery procedure of SERENADE. 
}

\subsection{Leader Election}\label{subsec:leader-election-embed}
\movedHL{
We explain this process on an arbitrary combinatorial cycle of $\sigma$, 
focusing on the actions of an arbitrary vertex $i$ that belongs to this cycle.  
We follow the standard practice~\cite{Perlman2000:leaderelection} of making the vertex with the
smallest identity (an integer between $1$ and $N$) on this cycle the leader.  
}\modifiedHL{
Recall that in the knowledge-discovery procedure, after each (say $k^{th}$) iteration,
vertex $i$ discovers $\sigma^{-2^{k}}(i)$ that is ``$2^k$ $\sigma$-hops away" from it on the cycle. 
More precisely, vertex $i$ learns $\phi^{(i)}_{k-}$, which contains the identities of the vertex $\sigma^{-2^k}(i)$, and 
the red and green weights of the walk 
$\sigma^{-2^k}(i)\leadsto i$.
}\movedHL{
Our goal is to augment this $k^{th}$ iteration to learn the vertex with the smallest identity on 
this walk $\sigma^{-2^k}(i)\!\leadsto\! i$, 
}\modifiedHL{
which we denote as $\calL(\sigma^{-2^k}(i)\!\leadsto\! i)$
and call 
the leader of the level-$k$ precinct right-ended at $i$.
}

\modifiedHL{
Like in the knowledge-discovery procedure, we explain this augmentation inductively. 
The case of $k\!=\!0$ ({\it i.e.}, the $0^{th}$ iteration) is as follows:  Each vertex $i$ considers 
the one with smaller identity between itself and $\sigma^{-1}(i)$ 
to be the leader of the level-0 precinct right-ended at $i$.

For the $k^{th}$ ($k\!\ge\!1$) iteration, each vertex $i$ only needs to augment the knowledge set it 
sends downstream to $i_D\!=\!\sigma^{2^{k-1}}(i)$ with $\mathcal{L}(\sigma^{(-2^{k-1})}(i)\!\leadsto\!i)$ (the leader of the level-$(k\!-\!1)$ precinct right-ended at $i$)
in~\autoref{serenade: send-down} of~\autoref{alg: serenade-general}. Meanwhile, it 
receives, from $i_U\!=\!\sigma^{(-2^{k-1})}(i)$, the vertex $2^{k\!-\!1}$ $\sigma$-hops upstream, 
$\mathcal{L}(\sigma^{(-2^{k-1})}(i_U)\!\leadsto\!i_U)$ (the leader of the level-$(k-1)$ precinct right-ended at $i_U$) 
in~\autoref{serenade: receive-up} of~\autoref{alg: serenade-general}, 
as $i_U$ also augments its knowledge set. In addition, each vertex $i$ also adds the following local computation 
in~\autoref{serenade: compute-knowledge-set-minus} of~\autoref{alg: serenade-general}.
\setlength{\belowdisplayskip}{0pt} \setlength{\belowdisplayshortskip}{0pt}
\setlength{\abovedisplayskip}{4pt} \setlength{\abovedisplayshortskip}{4pt}
\[\resizebox{0.95\hsize}{!}{
$\mathcal{L}(\sigma^{-2^{k}}(i) \leadsto i) \!\leftarrow\! \min\big{\{}\mathcal{L}(\sigma^{(-2^{k-1})}(i_U)\!\leadsto\!i_U), \mathcal{L}(\sigma^{(-2^{k-1})}(i)\!\leadsto\!i)\big{\}}$
}
\]
}

\modifiedHL{
The following lemma concerns the correctness of and the minimum number of iterations ({\it i.e.,} $1\!+\!\log_2N$) 
required by 
the above embedded leader election. This is also the reason why we choose to execute the knowledge-discovery 
procedure for $1\!+\!\log_2N$ iterations. Its proof is straightforward, we omit it in this paper.

\begin{lemma}\label{thm:leader-election}
Given any non-ouroboros cycle, each vertex $i$ belonging to it will learn, through the augmented knowledge-discovery procedure, 
the identity of the leader for the cycle, after at most $1\!+\!\log_2N$ iterations. 
Besides, there exists some non-ouroboros cycle such that some of its vertices need at least $1\!+\!\log_2N$ iterations to 
learn the identity of the leader.
\end{lemma}
}

\subsection{Distribute Leaders' Decisions}\label{subsec:leader-decision-notification}

\modifiedHL{
Once the leader of a non-ouroboros cycle is decided, through a distributed binary search (to be described in~\autoref{subsec:bs}), 
the leader will discover itself (through a non-empty walk). 
According to~\autoref{fact:disc}, the leader now can make the same matching decision as it would under SERENA. Then, 
the leader informs the switch controller of its decision on whether to 
choose the green or the red sub-matching, and the switch controller then broadcasts decisions of all leaders to 
the $N$ vertices. 
Since each vertex on a non-ouroboros cycle knows the identity of its leader by~\autoref{thm:leader-election},
it will follow the decision made by its leader in choosing between the red and the green sub-matchings. 
}

\movedHL{
The size of this broadcast, equal to the number of {\it non-ouroboros} cycles in $\sigma$, 
is small (with overwhelming probability).
}\modifiedHL{
For example, we will show in
\autoref{subsec:message-comp} 
that even when $N \!=\! 256$,  
the average 
number of non-ouroboros cycles is no more than $1.69$ and 
in more than $99\%$ of instances, there are no more than $4$ 
non-ouroboros cycles per time slot. 
}
\begin{algorithm}[tb]
\caption{Distributed binary search at vertex $i$.}
\label{alg: serenade-bsearch}
\SetKwRepeat{Do}{do}{while}
\SetKwProg{Proc}{Procedure}{}{end}
\SetKwFunction{algo}{\small BinarySearch}


\Proc{BinarySearch($i$, $k$, $w_g$, $w_r$)}{\label{bsearch: entrance}

    If $i$ (self) is $\mathcal{L}_0$ then halt\;
    \label{code: exit-check}


    \eIf{$\sigma^{(-2^{k-1})}(i)=\sleader_0$}
    {\label{code: endpoint-check}\label{code: middle-check}

        $w_g \leftarrow w_g - w_g\bigParenthes{\sigma^{(-2^{k-1})}(i) \leadsto i}$\;
        \label{code: green-weight-update}

        $w_r \leftarrow w_r - w_r\bigParenthes{\sigma^{(-2^{k-1})}(i) \leadsto i}$\;
        \label{code: red-weight-update}

        \algo{$\calL_0,k \!-\! 1,w_g,w_r$}\;
        \label{code: final-passing}

    }
    {

        \eIf{$\mathcal{L}_0=\calL(\sigma^{(-2^{k-1})}(i) \leadsto i$)} {\label{code: walk-leader-check}

            \algo{$i$, $k - 1$, $w_g$, $w_r$}\;
            \label{code: dummy-passing}
        }
        {


            $w_g \leftarrow w_g - w_g\bigParenthes{\sigma^{(-2^{k-1})}(i) \leadsto i}$\;
            \label{code: green-weight-update-}

            $w_r \leftarrow w_r - w_r\bigParenthes{\sigma^{(-2^{k-1})}(i) \leadsto i}$\;
            \label{code: red-weight-update-}

            \algo{$\sigma^{(-\!2^{k \!-\!1})}(i){,}k \!-\! 1{,}w_g{,}w_r$}\;
            \label{code: intermediate-passing}
        }
    }

} 
\end{algorithm}

\section{Distributed Binary Search Stage}\label{subsec:bs}\label{sec:bs}

\modifiedHL{
As mentioned above, only vertices on ouroboros cycles can make the exact same decisions as 
they would under SERENA, for vertices 
on non-ouroboros cycles, SERENADE needs an additional distributed binary search stage. In this section, we will describe the binary search stage 
focusing on an arbitrary non-ouroboros cycle.
}

\subsection{Distributed Binary Search}\label{subsubsec: b-search}\label{subsec: b-search}
\movedHL{
Without loss of generality, we assume that $\calL_0$ is the leader of, and $i$ a
vertex on, this non-ouroboros cycle. 
}\modifiedHL{
The objective of this distributed algorithm is 
to let its leader $\calL_0$ discovers itself 
twice by searching a repetition ({\it i.e.,} other than its first occurrence as 
the starting point of the walk) of $\calL_0$ 
along the walk $\mathcal{L}_0\!\leadsto\!\sigma^N(\mathcal{L}_0)$, the level-($\log_2 N$) precinct 
right-ended at $\sigma^N(\mathcal{L}_0)$;  this repetition must exist
because $N$, the length of the walk $\mathcal{L}_0 \leadsto \sigma^N(\mathcal{L}_0)$, is no smaller than
the length of this cycle.   
To this end, vertices on this non-ouroboros cycle perform a distributed binary search, guided by the
leadership information 
each vertex obtains through the leader
election. 
In the following, we describe the high-level ideas of this binary search algorithm, 
in which the detailed actions of a vertex $i$ are captured by~\autoref{alg: serenade-bsearch}. 
Unlike the knowledge-discovery procedure, during each iteration of the distributed binary search, only one 
vertex on this non-ouroboros cycle performs the search task, which we call {\it the search administrator}. 
}

\movedHL{
\medskip
\noindent
{\bf High-Level Ideas}. 
This binary search is initiated by the vertex
$\sigma^{N}(\calL_0)$, who learns ``who herself is" (\ie that herself is $\sigma^{N}(\calL_0)$) during the last iteration 
of the augmented knowledge-discovery procedure;  in other words, the initial {\it search administrator} is $\sigma^{N}(\calL_0)$.
The initial {\it search interval} is the entire walk $\calL_0\leadsto\sigma^{N}(\calL_0)$, also the level-$(\log_2 N)$ precinct right-ended at $\sigma^{N}(\calL_0)$.
}\modifiedHL{
The search administrator $\sigma^{N}(\calL_0)$ first checks whether itself or $\sigma^{N/2}(\calL_0)$, the middle point 
of the search interval, is a repetition of $\calL_0$. If so, the entire search mission is accomplished, so the 
search ends. Otherwise, it checks 
whether there is a repetition of $\calL_0$ in the right half of the search interval by checking whether
$\calL(\sigma^{N/2}(\calL_0)\leadsto\sigma^{N}(\calL_0))$ is equal to $\calL_0$; 
}\movedHL{
note the identity of $\calL(\sigma^{N/2}(\calL_0)\leadsto\sigma^{N}(\calL_0))$, 
the leader of the level-$(\log_2 (N) - 1)$ precinct right-ended at $\sigma^{N}(\calL_0)$,
is known to $\sigma^{N}(\calL_0)$, 
}\modifiedHL{
since it is one of the leadership information $\sigma^{N}(\calL_0)$ learns through the leader election.
}\movedHL{
If so, the same search administrator $\sigma^{N}(\calL_0)$ carries on this binary search in the right half of the search interval.  
Otherwise, the middle point of the search interval $\sigma^{N/2}(\calL_0)$ becomes the new search administrator and carries on this binary search in the left half. 
} 

\modifiedHL{
\medskip
\noindent
{\bf Pseudocode Explanation}. 
As mentioned above, the detailed actions of any search administrator $i$ are captured by~\autoref{alg: serenade-bsearch}. 
Initially $i$ is $\sigma^{N}(\calL_0)$ who assigns $\log_2 N$ to $k$, the $2^{nd}$ argument 
of~\autoref{alg: serenade-bsearch}, which indicates the search interval is $\sigma^{-2^k}\!\leadsto\!i$. To ensure that 
$\calL_0$ can discover itself, {\it i.e.,} learning the green and red weights of a non-empty walk from 
$\calL_0$ to itself, search administrator $i$ also maintains the weight information $w_g$, $w_r$ (the $3^{rd}$ and $4^{th}$ 
arguments of~\autoref{alg: serenade-bsearch}), 
which are the green and red weights 
of the walk from $\calL_0$ to $i$. Initially, search administrator $\sigma^N(\mathcal{L}_0)$ 
knows the green and red weights of the walk $\mathcal{L}_0\!\leadsto\!\sigma^N(\mathcal{L}_0)$, because 
they belong to the knowledge sets that $\sigma^N(\mathcal{L}_0)$ learns during the last iteration of the 
augmented knowledge-discovery procedure. We will not describe~\autoref{alg: serenade-bsearch} line-by-line, 
since the actions of search administrator $i$ are the same as those of the initial search administrator 
we described above, except that \autoref{alg: serenade-bsearch} details the operations for bookkeeping 
weight information.

The correctness of the distributed binary search and the number of iterations it requires, which 
are summarized in the following lemma, can be proved with mathematical induction. Here, we omit 
it for brevity. 
\begin{lemma}\label{lemma:bs}
Given any non-ouroboros cycle with a length of $\ell$, the distributed binary search enables the leader 
$\mathcal{L}_0$ of this cycle to discover itself twice with at most 
$\lceil \log_2\ell \rceil$ iterations.
\end{lemma}
}



\subsection{Complexity Analysis}\label{subsubsec:comp-bs}\label{subsec:comp-bs}

\modifiedHL{
In this section, we analyze the time and message complexities of the distributed binary search. 
}

\modifiedHL{
\medskip
\noindent
{\bf Time Complexity}. 
The time complexity ({\it i.e.,} number of iterations) of the distributed binary search 
is upper-bounded by $\lceil\log_2 \eta\rceil$ ($\le\! \log_2 N$) iterations, 
where $\eta$ is the length of the longest non-ouroboros cycle, since binary searches at different 
non-ouroboros cycles are performed simultaneously. 
Hence, it can be as large as $\log_2 N$ iterations, each of which contains roughly $3$ operations ($2$ additions and $1$ comparison), in the worst case. 
}



\modifiedHL{
\medskip
\noindent
{\bf Message Complexity}. 
As explained in~\autoref{subsubsec: b-search}, 
during the binary search, on each non-ouroboros cycle a message 
(to maintaining the weight information) 
is transmitted only when the search administrator moves from one vertex to another, 
so the message complexity of the binary search, in the worst-case, 
is at most $1$ message per vertex. Each message needs $\lceil\log_2\log_2N\rceil\!+\!C$ bits, where $C$ 
is the number of bits for encoding $w_r(\cdot)\!-\!w_g(\cdot)$. 
}
\section{Early Stop: O-SERENADE}
\label{subsec: o-serenade}

\modifiedHL{
In this section, we present an early-stop version of SERENADE to approximately emulate SERENA, 
without performing the distributed binary search, in which 
vertices on any non-ouroboros cycle make a
decision based on the (insufficient) information at hand after the augmented knowledge-discovery procedure. 
As will be shown in~\autoref{sec:evaluation}, 
O-SERENADE trades no degradation of delay performances 
for significant reduction in time complexities ({\it i.e.,} 
without performing the distributed binary search 
that has up to $\log_2N$ iterations). }

\medskip
\noindent
{\bf Decision Rule.}
\modifiedHL{Now we describe the decision rule of this early-stop version in an arbitrary non-ouroboros cycle. 
The decision rule is for the leader $\calL_0$ of this cycle to compare the green
and the red weights of the longest such walk 
$\calL_0\!\leadsto\!\sigma^N(\calL_0)$,
and pick, on behalf of the whole cycle, the green or the red sub-matching according to the outcome 
of this comparison;
}
note we cannot simply let every vertex $i$ on this cycle to pick the green or the red 
edge individually based on its local view of $w_r\big{(}i\leadsto\sigma^N(i)\big{)}$ 
vs. $w_g\big{(}i\leadsto\sigma^N(i)\big{)}$, since these local views can be inconsistent. 
\modifiedHL{
For example, for the leftmost cycle in~\autoref{fig: comb-cycles}, 
it is not hard to check vertex $2$ and vertex $3$ have inconsistent local views: 
$w_g\big{(}2\!\leadsto\!\sigma^{16}(2)\big{)}\!=\!120\!<\!134\!=\!w_r\big{(}2\!\leadsto\!\sigma^{16}(2)\big{)}$ 
(vertex $2$'s local view) and $w_g\big{(}3\!\leadsto\sigma^{16}(3)\!\big{)}\!=\!130\!>\!112\!=\!w_r\big{(}3\!\leadsto\sigma^{16}(3)\!\big{)}$ (vertex $3$'s local view). 
It is clear that O-SERENADE will pick the red sub-matching in this cycle based on the local view of its leader ({\it i.e.,} vertex $2$), which 
is different from the decision under SERENA.
}
This strategy is opportunistic as it does not hesitate to pick 
a sub-matching that appears to be larger, even though there is a small chance this appearance is incorrect.  
Therefore, we call it O-SERENADE. 
\modifiedHL{
Like SERENADE, this early-stop version also needs the switch controller to broadcast the decisions of all the leaders to the $N$ vertices.}

\medskip
\noindent
{\bf Rationale.}
The rationale behind the opportunistic strategy is as follows. 
When the weight difference between the red and the
green sub-matchings is small, 
it matters little which sub-matching
is picked.  When the difference is large, however, this strategy 
likely will further inflate the already large difference and hence result in the correct sub-matching being picked.




\section{Performance Evaluation}
\label{sec:evaluation}





\modifiedHL{
In this section, we evaluate, through simulations, 
the throughput and  
delay performances of O-SERENADE under various load conditions and traffic patterns 
to be specified 
in \autoref{subsec: sim-setting}; 
there is no need to evaluate the throughput and delay performances of SERENADE, 
which exactly emulates SERENA. 
{\bf Note that we have also evaluated the message complexity of SERENADE, and 
investigated how the mean delay performance of O-SERENADE scales with respect to N, the number of 
(input/output) ports; 
these results can be found in 
\autoref{app:more-sim}.}
}

\subsection{Simulation Setup}
\label{subsec: sim-setting}



\modifiedHL{
In all our simulations, 
the number of input/output ports $N$ is $64$, unless otherwise stated.  
}
To measure throughput and delay accurately, we assume each VOQ has an infinite buffer size and hence there is no packet drop at any input port. 
\modifiedHL{
Every simulation run lasts $30,000 \times N^2$ time slots.  
}
This duration is chosen so that every simulation run enters the steady state after a tiny fraction of this 
duration and stays there for the rest. The throughput and delay measurements are taken after the simulation run enters the steady state.





Like in \cite{GiacconePrabhakarShah2003SerenaJSAC}, we assume, in the following simulations, 
that the traffic arrival processes to different input ports are mutually independent, and each such arrival process is {\it i.i.d.}~Bernoulli (\ie at any given input port, 
a packet arrives with a constant probability $\rho \in (0,1)$ during each time slot).  
{\bf Note that we only use synthetic traffic (instead of that
derived from packet traces) because,
to the best of our knowledge, there is no meaningful way to 
combine packet traces into switch-wide traffic workloads.}  
The following $4$ standard types of load matrices (\ie traffic patterns) are used to generate the workloads of the switch: 
(I) \emph{Uniform}: packets arriving at any input port go
to each output port with probability $\frac{1}{N}$.
(II) \emph{Quasi-diagonal}: packets arriving at input port $i$ go to
output port $j \!=\! i$ with probability $\frac{1}{2}$ and go to any other output port
with probability $\frac{1}{2(N-1)}$.
(III) \emph{Log-diagonal}: packets arriving at input port $i$ go
to output port $j = i$ with probability $\frac{2^{(N-1)}}{2^N - 1}$ and
go to any other output port $j$ with probability equal $\frac{1}{2}$ of the
probability of output port $j - 1$ (note: output port $0$ equals output port $N$).
(IV) \emph{Diagonal}: packets arriving at input port $i$ go to
output port $j \!=\! i$ with probability $\frac{2}{3}$, or go to output port
$(i\, \text{mod} \, N) + 1$ with probability $\frac{1}{3}$.
%

The load matrices are listed in order of how skewed the traffic volumes to different output ports are: from uniform being the least skewed, to diagonal being the most skewed. 
Finally, we emphasize that, 
every non-zero diagonal element ({\it i.e.}, traffic from an input port $i$ and an output port $i$), in every traffic matrix we simulated on, is actually {\it switched} by the crossbar and consumes just as much 
switching resources per packet as other traffic matrix elements, 
\modifiedHL{
and never takes advantage of the ``local bypass'' (see \autoref{subsubsec: logical_bypass}) 
that may exist between the input port $i$ and the output port $i$. 
}
\subsection{Throughput Performance}\label{subsubsec:throughput-performance}\label{subsec:throughput-performance}
\modifiedHL{
Our simulation results show that O-SERENADE 
can achieve $100\%$ throughput under all $4$ load matrices and {\it i.i.d.} Bernoulli traffic arrivals: 
The VOQ lengths remain
stable under an offered load of $0.99$ in all these simulations.  
}



\subsection{Delay Performance}\label{subsubsec:delay-performance}\label{subsec:delay-performance}
\modifiedHL{
Now we shift our focus to the delay performance of O-SERENADE.
We compare its delay performance only with that of SERENA and MWM. 
We refer readers to \cite{GiacconePrabhakarShah2003SerenaJSAC} for 
comparisons between SERENA and some other crossbar scheduling algorithms 
such as iSLIP~\cite{McKeown99iSLIP} and iLQF~\cite{McKeown1995iLQF}.  
}

\begin{figure*}
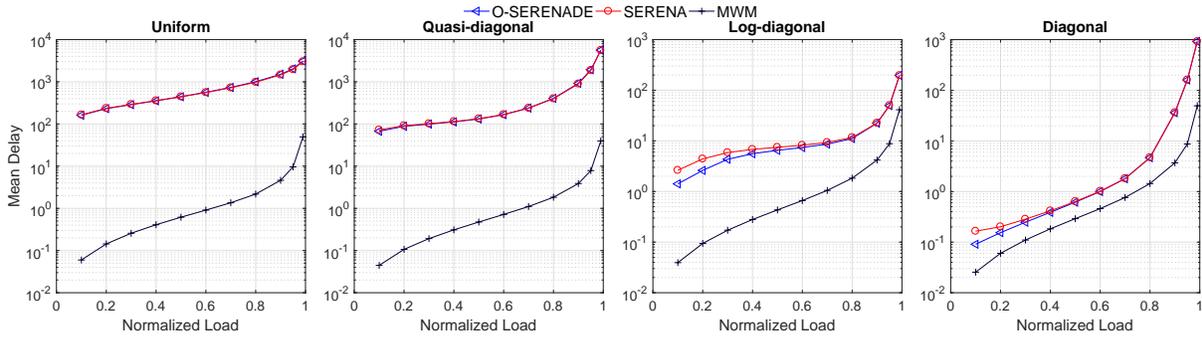

  \centering
  \MeanDelayvsLoadsO
  \caption{Mean delays of O-SERENADE, SERENA and MWM under
   the $4$ traffic load matrices.}
  \label{fig: delay-vs-loadsO}
\end{figure*}
\deletedHL{
C-SERENADE in general should perform worse (in mean delay) than SERENA, 
due to its conservative decision rule of sticking to the old (green) sub-matching on any non-ouroboros cycle,
even when it is not as heavy as the new (red) sub-matching.  
However, the ``damage" of being conservative 
to the delay performance is much larger
when the traffic is light -- and VOQ lengths are small -- for the following reason.  
The old sub-matching ``loses weight" at a high relative (to its current weight)
rate, and some of its VOQs (edges) could even become empty (\ie have weight $0$), whereas the new sub-matching 
can ``gain weight" at a high relative rate.  
This damage becomes tiny when the traffic is very heavy -- and VOQ lengths are long -- because once C-SERENADE 
settles in a very heavy matching, the matching (and its sub-matchings) will remain very heavy, relative to both its current weight and 
those of other ``rising star" (in weight) matchings, for quite a while.

When the traffic load is light (say $0.1$), the arrival graph has very few edges, 
which means that most edges in the derived matching ({\it i.e.,} $S_r$) are 
added though the random population. Its effective size ({\it i.e.,} number 
of edges 
whose weight is not zero) is highly likely to be small when the switching system 
starts, and so is that of the merged matching. 
Hence, queues build up. As a result of that, the effective size of $S_r$ 
is more likely to be large so that the merged matching is more likely to 
get improved. Once the merged matching improves to such a level that its 
effective size is larger than the average total arrival rate, queues would 
drop. Hence the merged matching also gets harder to be improved so that 
its effective size would eventually decrease to a level 
lower than the arrival rate. 
Then, queues build up again. When these queue building up and dropping 
get counterbalanced, the average queue size gets stationary, so does the 
average mean delay. 

As the traffic load gets higher, the arrival graph has more edges and the 
effective size of the derived matching is more likely to be large. 
However, it is also more difficult to counterbalance the queue building up 
because of higher arrival rate. Thus, generally we would expect the average 
queue size increases as the traffic load gets higher for SERENADE. However, 
in C-SERENADE, the conservative decision introduces a new ``force'' that 
prevents the merged matching getting improved. That is, we have described in 
\autoref{subsec: c-serenade}, the conservative decision only accepts the 
sub-matchings of $S_r$, which have larger weight and belong to the 
Ouroboros cycles. Therefore, the average mean delay drops as the traffic 
load increases when the load is not high enough (say $<0.6$).

{\bf Another version}.

in a switching system using SERENADE as the scheduling 
algorithm is determined by the game between the arrivals because of the 
offered load and the departures based on the schedule produced by SERENADE. 
As the offered load gets higher, 
the arrival matching is more likely to have more edges, hence more 
likely to result a merged matching with higher weights. Note that higher 
weights are not equivalent to higher departure rates, but it means that 
SERENADE is more likely to find a better matching before the effective 
size of the matching ({\it i.e.,} number of edges that have positive weights), 
which is equivalent to the departure rate, decreases too much. However, 
the queue also builds up faster because of higher offered load, so 
SERENADE requires more packets in the switching system, {\it i.e.,} a 
higher average queue length in the stationary stage, 
to counterbalance the faster arrivals. For C-SERENADE, since the conservative 
decision rule prevents improvements in any non-ouroboros cycle, it 
requires a higher average queue size. Especially, when the offered load 
is light (say $0.1$). Because of the complicated interactions between 
those ``forces'', the average mean delay of C-SERENADE first 
decreases and then increases as the offered load increases. 
}

\medskip
\noindent
{\bf O-SERENADE vs. SERENA.} 
~\autoref{fig: delay-vs-loadsO} 
shows the mean delays of the three algorithms under the $4$ traffic load matrices above respectively.
Each subfigure shows how the mean delays (on a {\it log scale} along the y-axis) vary with different offered loads (along the x-axis).
\autoref{fig: delay-vs-loadsO} shows that overall O-SERENADE and SERENA perform similarly under all $4$ traffic load matrices and all load factors. 
Upon observing these simulation results, our interpretation was that the decisions made by O-SERENADE 
agree with the ground truth (\ie which sub-matching is indeed heavier on a non-ouroboros cycle) most of time.  This interpretation 
was later confirmed by further simulations:  They agree in between $90.57\%$ and $99.99\%$ of the instances.  

\modifiedHL{
Perhaps surprisingly, \autoref{fig: delay-vs-loadsO} also
shows that O-SERENADE performs slightly better than SERENA when the 
traffic load is low (say $<0.4$) under log-diagonal and diagonal traffic load matrices.  
}
Our interpretation of this observation is as follows.
It is not hard to verify that decisions made by O-SERENADE can disagree with the ground truth, with a non-negligible probability, 
only when the total green and the total red weights of a non-ouroboros cycle are very close to one another.  However, in such cases,
picking the wrong sub-matchings (\ie disagreeing with the ground truth) causes almost no damages.  Furthermore, we speculate that 
it may even help O-SERENADE jump out of a local maximum (\ie have the effect of simulated annealing) and 
converge more quickly to a near-optimal matching (in terms of weight), thus resulting in even better delay performance.

\ifExtension

\subsection{Bursty Arrivals}
\label{subsubsec: bursty-arrivals}

\begin{figure*}
  \centering
  \subfigure[Under the offered load of 0.6.]{
    \centering
    \MeanDelayvsBurstSizeModerate
  }\\
  \subfigure[Under the offered load of 0.95.]{
    \centering
    \MeanDelayvsBurstSize
  }
  \caption{Mean delays of SC-SERENADE, SO-SERENADE and SERENA under the 
  bursty arrivals with the $4$ traffic load matrices.}
  \label{fig: Mean-Delay-Burst}
\end{figure*}

\addedHL{In real networks, packet arrivals are likely to be bursty.
In this section, we evaluate the performances of SC-SERENADE, SO-SERENADE, and SERENA under heavy bursty traffic, generated by
a two-state ON-OFF arrival
process described in~\cite{GiacconePrabhakarShah2003SerenaJSAC}.
The durations of each ON (burst) stage and OFF (no burst) stage are
geometrically distributed: the probabilities that the ON and OFF states last for $t \ge 0$ time slots are given by
\begin{equation*}
P_{ON}(t) = p(1-p)^t \text{ and } P_{OFF}(t) = q(1-q)^t,
\end{equation*}
with the parameters $p, q \in (0,1)$ respectively. As such, the average duration of the
ON and OFF states are $(1-p)/p$ and $(1-q)/q$ time slots
respectively.

In an OFF state, an incoming packet's destination (\ie output
port) is generated according to the corresponding load matrix. In an
ON state, all incoming packet arrivals to an input port would be destined
to the same output port, thus simulating a burst of packet
arrivals. By controlling $p$, we can control the desired
average burst size while by adjusting $q$,
we can control the load of the traffic.

We have evaluated the mean delay 
performances of SC-SERENADE, SO-SERENADE, and SERENA, with the average burst size ranging from $1$ to $512$ packets, under heavy offered loads up to 0.95.  The simulation results under a moderate offered load of $0.6$ and a heavy
offered load of $0.95$, plotted 
in \autoref{subfig:moderate-bs} and \autoref{subfig:heavey-bs} respectively, show that SC-SERENADE and SO-SERENADE have almost the same mean 
delays as SERENA under the $4$ traffic load matrices with all the burst sizes, except 
that C-SERENADE has slightly longer delay at small burst size (say $<4$) under 
moderate traffic loads. 
Similar observations can be made about the mean delay performances of the three algorithms under other moderate and heavy offered loads.
To summarize, our simulation studies show conclusively that C-SERENADE and O-SERENADE, as well as their stabilized variants
SC-SERENADE and SO-SERENADE, are able to handle bursty traffic as well as SERENA.}






\fi 

\section{Related Work}
\label{sec: related-work}
In the interest of space, we provide only a brief survey of the prior art that is directly related to our work.
Since SERENADE parallelizes SERENA, which computes approximate Maximum Weight Matching (MWM), 
we focus mostly on the following two categories: 
(1) parallel or distributed algorithms for exact or approximate MWM computation with applications to 
crossbar scheduling (in \autoref{subsec: sched-alg-crossbar}) and 
(2) distributed matching algorithms with applications to transmission scheduling in wireless networks (in \autoref{subsec: sched-alg-wireless}).  
In particular, we will keep to a minimum the comparisons between SERENA and other sequential crossbar scheduling algorithms proposed before SERENA, 
\modifiedHL{
of which a thorough survey was provided in~\cite{GiacconePrabhakarShah2003SerenaJSAC}.
}
%
%

\modifiedHL{
A few {\it sequential} crossbar scheduling algorithms were proposed~\cite{Gupta09Node,Ye10Variable}.
}
However, none of these algorithms beats SERENA in both (delay and throughput) performance 
and computational complexity under the standard problem setting (\textit{e.g.,} fixed packet size).  
A template that can be instantiated into a family of throughput-optimal algorithms
for scheduling crossbar or wireless transmission, and a unified
framework for proving the throughput-optimality of all these algorithms were proposed in~\cite{ShinSuk2014OraclearXiv,ShinSuk2014Oracle}.
\modifiedHL{
The only crossbar scheduling algorithm that results from this template is the instantiation of a 
BP-based MWM algorithm~\cite{BayatiShahSharma2008MWMMaxProduct},
}
which has a message and time complexity of $O(N)$ per port. 
In~\cite{Hu2016IterSched}, an efficient distributed iterative algorithm, called 
RR/LQF, was proposed.  Although its computational complexity can be as low as one 
iteration (but at a cost), and its message complexity as low as one bit per port, it requires a crossbar speedup of $2 - 1/N$ to achieve $100\%$ throughput,  
if it runs $N$ iterations of the algorithm, where $N$ is the number of 
input/output ports. 
Recently, an ``add-on" algorithm called Queue-Proportional Sampling (QPS) was proposed in~\cite{GongTuneLiuEtAl2017QPS}
that can be used to augment, and boost the delay performance of, SERENA~\cite{GiacconePrabhakarShah2003SerenaJSAC}. 
However, the resulting QPS-SERENA has the same $O(N)$ time complexity as SERENA.

In all algorithms above, a matching decision is made every time slot.
An alternative type of algorithms is frame-based 
~\cite{Aggarwal2003EdgeColoring,Neely2007FrameBased,Li2008framebased,Wang2016ParallelEdgeColoring,Wang2018ParallelEdgeColoring},  
in which multiple (say $K$) consecutive time slots are grouped as 
a frame. 
These $K$ matching decisions in a frame are batch-computed, 
which usually has lower time complexity than 
$K$ independent matching computations. 
However, 
\modifiedHL{
since $K$ is usually quite large ({\it e.g.,} $K\!=\! O(\log N)$),
} 
and a packet arriving at 
the beginning of a frame has to wait till at least the beginning of the next frame to be switched, frame-based 
scheduling generally results in higher queueing delays. 

\subsection{Parallel/Distributed MWM Algorithms}\label{subsec: sched-alg-crossbar}

As mentioned earlier, 
\modifiedHL{
MWM is the ideal crossbar scheduling policy in the sense that it can achieve 100\% throughput and 
that it has delay performance conjectured to be optimal,
}
but its most efficient 
implementation~\cite{EdmondsKarp1972}
has a prohibitively high computational complexity of $O(N^3)$.  
This dilemma has motivated
the development of a few parallel or distributed algorithms that, by distributing 
this computational cost across multiple processors (nodes), bring down the per-node 
computational complexity. 

The most representative among them are \cite{Fayyazi04ParallelMWM,BayatiPrabhakarShahEtAl2007Iterative,BayatiShahSharma2008MWMMaxProduct,AtallaCudaGiacconeEtAl2013BPAssist}.
A parallel algorithm with a sub-linear per-node computational complexity of $O(\sqrt{N}\log^2 N)$ was proposed 
in~\cite{Fayyazi04ParallelMWM} for computing MWM exactly in a bipartite graph.  However, this algorithm requires the use of 
$O(N^3)$ processors.   Another two~\cite{BayatiPrabhakarShahEtAl2007Iterative,BayatiShahSharma2008MWMMaxProduct} belong to
the family of distributed iterative algorithms based on belief-propagation (BP). In this family, 
the input ports engage in multiple iterations of message 
exchanges with the output ports
to learn enough information about the lengths of all $N^2$ VOQs 
so that each input port
can decide on a distinct output port to match with. The resulting 
matching either is, or is close to, the MWM.  Note that the BP-based 
algorithms are simply parallel algorithms to compute the MWM: 
the total amount of computation, or the total number of messages 
needed to be exchanged, is still $O(N^3)$, but is distributed 
evenly across the input and the output ports (\ie $O(N^2)$ work for each 
input/output port).
It was shown in~\cite{AtallaCudaGiacconeEtAl2013BPAssist} that BP can also be used to boost the 
performance of other (non-BP-based) distributed iterative algorithms
such as iLQF~\cite{McKeown1995iLQF}.
However, the ``BP assistance" part alone has a total computational 
complexity of $O(N^2)$, or $O(N)$ per port.

\subsection{Wireless Transmission Scheduling}
\label{subsec: sched-alg-wireless}
Transmission scheduling in wireless networks with primary interference constraints~\cite{ModianoShahZussman2006DistSched} shares a common
algorithmic problem with crossbar scheduling:  to compute a good matching for each ``time slot".
The matching computation in the former case is however more challenging, since it needs to be 
performed over a general graph that is not necessarily bipartite.
Several wireless transmission scheduling solutions were proposed in the literature
\cite{Lin2005DistSched, ModianoShahZussman2006DistSched,Chen2006DistSched,Chaporkar2008DistSched, Gupta2009DistSched,Ji2013DistSched} 
that are based on distributed computation of matchings in a general graph.

Most of these solutions tackle the underlying distributed matching computation problem using an adaptation/extension of 
either \cite{IsraelItai1986DistMaximalMatching} (used in~\cite{Gupta2009DistSched,Chaporkar2008DistSched,Ji2013DistSched}), 
or \cite{Hoepman2004DistAppMWM} (used in~\cite{Chen2006DistSched,Lin2005DistSched}).  In~\cite{IsraelItai1986DistMaximalMatching}, a parallel randomized algorithm was proposed that outputs a maximal matching with expected runtime $O(\log |E|)$, where $|E|$ is the number of edges in the graph.  This computational 
complexity, translated into our crossbar scheduling context, is $O(\log N)$. 
However, 
maximal matching algorithms are known to only guarantee at least 
$50\%$ throughput~\cite{McKeownMekkittikulAnantharamEtAl1999}. The work of Hoepman~\cite{Hoepman2004DistAppMWM} converts an earlier sequential algorithm for computing approximate MWM~\cite{Preis1999AppMWM} to a distributed 
one. 
However, the distributed algorithm in~\cite{Hoepman2004DistAppMWM}, like its sequential version~\cite{Preis1999AppMWM}, can only guarantee to find a matching whose weight is at least half of that of the MWM, and hence can only 
guarantee at least $50\%$ throughput also. 
\modifiedHL{
In comparison, 
SERENADE guarantees $100\%$ throughput, 
just like its sequential version SERENA.
}

The only exception, to distributed matching algorithms being based on either~\cite{IsraelItai1986DistMaximalMatching} or~\cite{Hoepman2004DistAppMWM}, is~\cite{ModianoShahZussman2006DistSched}, in which the scheduling algorithm, called MIX, is 
a distributed version of the MERGE 
procedure in SERENA, albeit in the wireless networking context. 
The objective of MIX is to compute an approximate MWM 
for simultaneous non-interfering wireless transmissions of packets, 
where the weight of a directed edge (say a wireless link from a node $X$ to a 
node $Y$) is 
the length of the VOQ at $X$ for packets destined for $Y$, in the SERENA manner:  MERGE 
the matching used in the previous time slot 
with a new random matching.  Unlike in SERENA, however, neither matching has to be full and the connectivity topology is generally 
not bipartite in a wireless network, and hence the graph resulting from the union of the two matchings can contain both cycles and paths. 

\modifiedHL{
MIX has three variants.  
}
As we 
will explain in \autoref{subsec: serenade-vs-gossiping} 
in details, all three variants compute the total -- or equivalently the average -- green 
and red weights of each cycle or path either by 
linearly traversing the cycle or path, or via a gossip algorithm~\cite{BoydGhoshPrabhakarEtAl2005gossip};
they all try to mimic SERENA in a wireless network and have a time complexity at least $O(N)$, as compared to $O(\log N)$ for SERENADE.  
To summarize, they are clearly all ``wireless SERENA", not ``wireless SERENADE''.

\section{Conclusion}
\label{sec: conclusion}


\modifiedHL{
In this paper, we propose SERENADE, 
a parallel iterative algorithm that can provably, 
with a time complexity of only $O(\log N)$ per port, 
exactly emulate
SERENA, a centralized algorithm with $O(N)$ time complexity. 
We also propose an early-stop version of SERENADE, called O-SERENADE, which 
only approximately emulates SERENA. 
Through extensive simulations, we demonstrate that 
O-SERENADE can achieve $100\%$ throughput. 
We also demonstrate that O-SERENADE has delay performances either similar as or better than those of SERENA, under various load conditions and 
traffic patterns.
}

\bibliographystyle{hIEEEtran}
\bibliography{bibs/serenade-references,bibs/lyapunov_fluid}

\appendices
\section{Parallelized Population}
\label{subsec: parallel-population}



As explained in \autoref{sec: serena-and-merge}, 
the new random matching $A'(t)$ derived from the {\it arrival graph}, which is 
in general a partial matching, has to be populated into a full matching $R(t)$ before it can be 
merged with $S(t-1)$, 
the matching used in the previous time slot.  
SERENADE parallelizes this POPULATE procedure, \ie the round-robin pairing of unmatched input ports in $R(t)$ with unmatched output ports in $R(t)$, 
so that the computational complexity
for each input port is
$O(\log N)$, as follows. 

Suppose that each unmatched port (input port or output port) knows its own ranking, 
{\it i.e.,} the number of unmatched ports 
up to itself (including itself) from the first one (we will show later how each unmatched 
port can obtain its own ranking). Then, each unmatched input 
port needs to obtain the identity of the unmatched output port with the same 
ranking. This can be done via $3$ message exchanges as follows. 
Each pair of unmatched input and output ports ``exchange'' 
their identities through a ``broker''. More precisely, 
the $j^{th}$ unmatched input port ({\it i.e.,} unmatched input port with 
ranking $j$) first sends its identity to input port $j$ ({\it i.e.,} the 
broker). 
Then, the $j^{th}$ unmatched output port also sends its identity to input port $j$ ({\it i.e.,} the 
broker). 
Finally, input port $j$ ({\it i.e.,} the 
broker) sends the identity of 
the output port with ranking $j$ to the input port (with ranking $j$). 
Thus, the input port learns the identity of the corresponding output 
port. 
Note that, since 
every pair of unmatched input port and output port 
has its unique ranking, thus they would have different ``brokers''. Therefore, all pairs can 
simultaneously exchange their messages without causing any congestion ({\it i.e.,} a port sending or 
receiving too many messages). 

It remains to parallelize the computation of ranking each port (input port or output port). 
This problem can be reduced to 
the parallel prefix sum problem~\cite{Edelman2004ParallelPrefix} as follows. Here, we 
will only show how to compute the rankings of input ports in parallel; 
that for output ports is identical. 
Let $B[1..N]$ be a bitmap that indicates whether input port $i$ is unmatched (when $B[i]=1$) or not
(when $B[i]=0$).  
Note that, this bitmap is distributed, that is, each input port $i$ only has a single bit $B[i]$. 
For $i = 1,2,\cdots,N$, denote as $r_i$ the ranking of 
input port $i$.  It is clear that   
$r_i=\sum_{k=1}^{i}B[k]$, for any $1\le i \le N$.
In other words, the $N$ terms $r_1$, $r_2$, $\cdots$, $r_N$ are the prefix sums 
of the $N$ terms $B[1]$, $B[2]$, $\cdots$, $B[N]$. Using the Ladner-Fischer 
parallel prefix-sum algorithm~\cite{Ladner1980PrefixSum}, we can obtain 
these $N$ prefix sums $r_1$, $r_2$, $\cdots$, $r_N$ in $O(\log N)$ time (per port) using $2N$ processors (one at each input or output port). 

\section{Proofs}

\subsection{Proof of \autoref{lemma:disc}}\label{app:proof-lemma-01}
We need only to consider the following two cases.
\begin{enumerate}[wide, labelwidth=!, labelindent=0pt]
    \item {\bf The two walks are in the same ``rotational" direction}. Without loss of generality, we assume the two walks are 
    $i\!\leadsto\! \sigma^{\beta}(i)$ and $i\!\leadsto\! \sigma^{\gamma}(i)$ respectively, where 
    $0\! \le\! \beta \!<\! \gamma$, 
    and $\sigma^\beta(i)\!=\!\sigma^\gamma(i)\!=\!j$. By applying the operator $\sigma^{-\beta}$ to both sides 
    of the equation $\sigma^\beta(i)\!=\!\sigma^\gamma(i)$, we have $i\!=\!\sigma^{\gamma\!-\!\beta}(i)$. Hence, 
    $\ell$ divides ($\gamma\!-\!\beta$), where $\ell$ is the length of the cycle to which vertices $i,j$ belong. 
    Suppose $\kappa\ell\!=\!\gamma\!-\!\beta$, where $\kappa\!>\!0$ is an integer. 
    Then, we have the ($\gamma\!-\!\beta$)-edge-long 
    walk $\sigma^{\beta}(i)\!\leadsto\!\sigma^{\gamma}(i)$ coils around the cycle (of length $\ell$) exactly 
    $\kappa$ times, and so the green weight $w_g\big{(}\sigma^{\beta}(i)\!\leadsto\!\sigma^{\gamma}(i)\big{)}$ 
    (or red weight) is $\kappa$ times of that of the cycle. Since vertex $i$ can obtain the green weight (or the red weight) 
    of the walk $\sigma^{\beta}(i)\!\leadsto\!\sigma^{\gamma}(i)$ via subtracting $w_g(i\!\leadsto\!\sigma^\beta(i))$ from 
    $w_g(i\!\leadsto\!\sigma^\gamma(i))$, {\it i.e.,}
    \setlength{\belowdisplayskip}{0pt} \setlength{\belowdisplayshortskip}{0pt}
    \setlength{\abovedisplayskip}{4pt} \setlength{\abovedisplayshortskip}{4pt}
    \begin{equation}\label{eq:comp-weights}\resizebox{0.9\hsize}{!}{
    $\begin{cases}
      w_g\bigParenthes{\sigma^{\beta}(i)\!\leadsto\!\sigma^{\gamma}(i)} &\leftarrow
    w_g\bigParenthes{i\!\leadsto\!\sigma^\gamma(i)} - w_g\bigParenthes{i\!\leadsto\!\sigma^\beta(i)}\\
    w_r\bigParenthes{\sigma^{\beta}(i)\!\leadsto\!\sigma^{\gamma}(i)} &\leftarrow
    w_r\bigParenthes{i\!\leadsto\!\sigma^\gamma(i)} - w_r\bigParenthes{i\!\leadsto\!\sigma^\beta(i)}
    \end{cases}$
    }\end{equation}
    it knows whether $w_g(i\!\leadsto\!\sigma^\ell(i))$ (the green weight of the cycle) 
    or $w_r(i\!\leadsto\!\sigma^\ell(i))$ (the red weight of the cycle) is larger.
    \item {\bf The two walks are in opposite directions}. Without loss of generality, we assume the two walks are 
    $\sigma^{-\beta}(i)\!\leadsto\! i$ and $i\!\leadsto\! \sigma^{\gamma}(i)$ respectively where $\beta,\gamma \!>\! 0$, and 
    $\sigma^{-\beta}(i)\!=\!\sigma^\gamma(i)\!=\!j$. By applying the operator $\sigma^{\beta}$ to both sides 
    of the equation $\sigma^{-\beta}(i)\!=\!\sigma^\gamma(i)$, we have $i\!=\!\sigma^{\gamma+\beta}(i)$. So 
    $\ell$ divides ($\gamma\!+\!\beta$), the rest reasoning is the same as before.
\end{enumerate}

\subsection{Proof of~\autoref{lemma:all-or-nothing}}\label{app:proof-of-lemma-02}
Suppose that vertex $i$ discovers vertex $j$ after the $k_1^{th}$ and $k_2^{th}$ iteration respectively. 
Then we have $\sigma^{m_1}(i)\!=\!\sigma^{m_2}(i)\!=\!j$ where $m_1\!=\!2^{k_1}$ if $i$ discovers $j$ 
through~\autoref{serenade: compute-knowledge-set} during the $k_1^{th}$ iteration, otherwise $m_1\!=\!-2^{k_1}$. 
Similarly, $m_2\!=\!\pm2^{k_2}$. Thus, we have $i\!=\!\sigma^{m_2-m_1}(i)$. Therefore, there exists 
some positive integer $\kappa$ such that $\abs{m_2-m_1} \!=\!\kappa \ell$, where $\ell$ is the length of 
the cycle. 

For any other vertex $x$ on the same cycle, we have $x\!=\!\sigma^{m_2-m_1}(x)$. Thus, 
$\sigma^{m_2}(x)\!=\!\sigma^{m_1}(x)\triangleq y$. Therefore, $x$ also discovers $y$ twice.

\subsection{Proof of~\autoref{lemma:ouroboros-lemma}}\label{app:proof-of-lemma-03}

\begin{figure}[!htbp]
\centering
\includegraphics[width=0.45\textwidth]{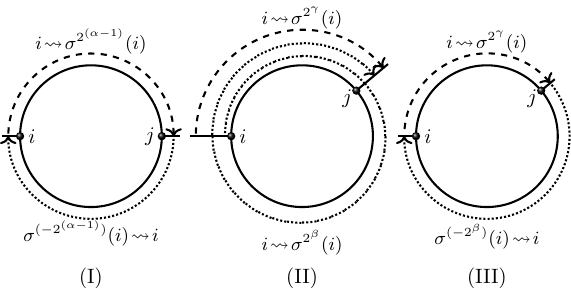}
\caption{Illustration of three cases corresponding to cycle lengths belonging to the three forms of the ouroboros numbers.}
\label{fig:ouroboros-numbers}
\end{figure}

Here we only show the proof sketch for brevity. 
It is not hard to show that if the cycle, to which vertex $i$ belongs, has a length of an ouroboros number, {\it i.e.,} 
a divisor of numbers  
in the three forms defined in~\autoref{def: ouroboros-cycle}, 
then vertex $i$ will discover a vertex $j$ on the same cycle twice via one of the three 
cases illustrated in~\autoref{fig:ouroboros-numbers}. For example, if the cycle length 
$\ell$ is a divisor of form (I) defined in~\autoref{def: ouroboros-cycle}, then it has to be a power of 
$2$. Without loss of generality, we assume $\ell\!=\!2^\alpha$, where nonnegative integer $\alpha\!\leq\!\log_2 N$. 
In the cases of $\alpha \ge 1$, it is clear that, in the $(\alpha \!-\! 1)^{th}$ iteration, 
vertex $i$ discovers $\sigma^{2^{\alpha\!-\!1}}(i)$ and $\sigma^{(-2^{\alpha\!-\!1})}(i)$, 
which turn out to be the same vertex. The case of $\alpha\!=\!0$ is slightly different: After the $0^{th}$ iteration, 
vertex $i$ discovers $\sigma(i)$ and 
$\sigma^{-1}(i)$, which are both equal to $i$. 
Therefore, (I) shown in~\autoref{fig:ouroboros-numbers} happens. 
Similarly, 
we can show that if the cycle length $\ell$ is a divisor of a number in form (II) (or (III)), then (II) (or (III)) 
shown in~\autoref{fig:ouroboros-numbers} happens. 

Note that for (II) and (III) in~\autoref{fig:ouroboros-numbers}, 
both the two walks can coil around the cycle for one or more times, and the directions of the two 
walks can be reversed. For example, for (III) in~\autoref{fig:ouroboros-numbers}, the two walks could also 
be $\sigma^{-2^\gamma}(i)\!\leadsto\!i$ and $i\!\leadsto\!\sigma^{2^\beta}(i)$, and both of them could 
coil around the cycle for one or more times, {\it i.e.,} they are longer than the cycle.

\section{Why Not Use More Than $1+\log_2 N$ Iterations?}\label{app:why-not-more}

Fix a vertex $i$.  Note that in the knowledge-discovery procedure, each iteration results in two new vertices being discovered by vertex $i$
and hence increases the chance of a vertex being discovered twice by $i$.  Hence, if we run more than $1\!+\!\log_2 N$ iterations, then vertex $i$ may discover a vertex twice
even if it is on a non-ouroboros cycle (as defined in~\autoref{def: ouroboros-cycle}).  
In other words, with additional iterations, some 
non-ouroboros numbers may become ``effective ouroboros numbers".    Readers may wonder if we can do away with the distributed binary search simply by running a little more iterations (say $0.5\log_2N$ more iterations).
Unfortunately, as shown in~\autoref{tab:max-iter}, 
there exists some numbers (cycle lengths) that are ``hardcore non-ouroboros" in the sense a vertex $i$ on a cycle of such a length $\ell$ needs to 
run exactly $\lceil \ell/2\rceil$ iterations to discover a vertex twice.
In fact, it is a long-standing open problem in mathematics whether there exists infinite number of what we call ``hardcore non-ouroboros" numbers here.  
More precisely, it is a special case of the Artin's Conjecture~\cite{stein2008elementary}, which, if put into
our context, asks whether there are infinitely many prime numbers $p$ such that, it takes a vertex $i$ on a cycle of length $p$
exactly $\lceil p/2\rceil$ iterations to discover a vertex on the same cycle twice.

\begin{table}[!htbp]
\centering
\caption{Examples of ``hardcore non-ouroboros'' numbers.}\label{tab:max-iter}
\begin{tabular}{@{}rccccc@{}}
\toprule
$\ell$ & 61 & 131 & 239 & 509 & 1019 \\\midrule
Iterations & 31 & 66 & 120 & 255 & 510\\\bottomrule
\end{tabular}
\end{table}

\section{SERENADE vs. MIX}
\label{subsec: serenade-vs-gossiping}

In this section, we describe the three variants of MIX~\cite{ModianoShahZussman2006DistSched} in detail.
The first variant, which is centralized and idealized, computes the total green and red weights of each cycle or path by ``linearly" traversing the cycle or path.  
Hence it has a time complexity of $O(N)$, where $N$ is the number of nodes in a wireless network. 
This idealized variant is however impractical because it requires the complete knowledge of the connectivity topology of the wireless network.

The second variant removes this infeasible requirement and hence is practical.  It estimates and compares 
the {\it average} green and red weights of each cycle or path (equivalent to comparing the total green and red weights) using a 
synchronous iterative gossip algorithm proposed in~\cite{BoydGhoshPrabhakarEtAl2005gossip}.  In this gossip algorithm, each node
(say $X$) is assigned a green (or red) weight that is equal to the weight of the edge that uses $X$ as an endpoint and belongs to 
the matching used in the previous time slot (or in the new random matching);   
in each iteration, each node attempts to pair with a random neighbor and, if this attempt is successful, both nodes will be 
assigned the same red (or green) weight equal to the average of their current red (or green) weights.  
The time complexity of each MERGE is $O(l^2 N\log N)$, since this gossip algorithm requires $O(l^2 N\log N)$ iterations~\cite{ModianoShahZussman2006DistSched} for the average red (or green) weight estimate to be 
close to the actual average with high probability.  Here $l$ is the length of the longest path or cycle.

The third (practical) variant, also a gossip-based algorithm, employs the aforementioned ``idempotent trick" (see \autoref{subsubsec: idempotent}) 
to estimate and compare the
{\it total} green and red weights of each cycle or path.  This idempotent trick reduces the convergence time 
(towards the actual total weights) to $O(l)$ iterations, but as mentioned earlier 
requires each pair of neighbors to exchange a large number ($O(N\log N)$ to be exact) of exponential random variables during each message exchange.
Since $l$ is usually $O(N)$ in a random graph, the time complexity of this algorithm can be considered $O(N)$.

\section{An Idempotent Trick}
\label{subsubsec: idempotent}



As mentioned in 
\autoref{subsec: serenade-vs-gossiping}, 
there is an alternative solution to 
the consistency problem that does not require leader election, using a standard ``idempotent trick" 
that was used in~\cite{ModianoShahZussman2006DistSched} to solve a similar problem.   To motivate this trick, we zoom in on the example 
shown in 
\autoref{fig: comb-cycles}.  
Both the 
consistency problem and the absolute correctness problem above can be attributed to the fact that the (green or red) weights of some edges are accounted for ({\it i.e.}, added to the total)
$\kappa$ times, while those of others $\kappa - 1$ times.   For example, in 
{\bf $3\leadsto \sigma^{16}(3)$, the (green or red) weights of edges $(3,4)$, $(4,14)$, and {\it etc} are accounted for
$2$ times, while those of edges $(2,8)$, $(8,5)$, and {\it etc} $1$ times.} 
Since the ``+" operator is not idempotent (so adding a number to a counter $\kappa$ times is not the same as adding it $\kappa - 1$ times), the total (green or red) weight of the walk obtained this way does not perfectly track that of 
the cycle.  

The ``idempotent trick" is to use, instead of the ``+" operator, a different and idempotent operator $MIN$ to arrive at an estimation of the total green (or red) weight;  the $MIN$ is idempotent in the sense the minimum of a multi-set (of real numbers) $M$ is the same as that of the set of distinct 
values in $M$.
The idempotent trick works, in this O-SERENADE context, for a set of edges $e_1$, $e_2$, ..., $e_Z$ that comprise a non-ouroboros cycle 
with green weights $w_1$, $w_2$, ..., $w_Z$ respectively, as follows;  the trick works in the same way for the red weights.
Each edge $e_\zeta$ ``modulates" its green weight $w_\zeta$ onto an exponential random variable $X_\zeta$ with distribution $F(x)=1-e^{-x/w_\zeta}$ (for $x > 0$) so that $E[X_\zeta] 
= w_\zeta$.  Then the green weight of every walk $W$ on this non-ouroboros cycle can be encoded as $MIN\{X_\zeta | e_\zeta \in W\}$.  It is not hard to show
that we can compute this MIN encoding of every $2^d$-edge-long walk $i\leadsto\sigma^{2^d}(i)$ by SERENADE-common in the same inductive way we compute 
the ``+" encoding. For example, under the MIN encoding, \autoref{serenade: send-up} in \autoref{alg: serenade-general} becomes
``Send to $i_U$ the value $MIN\{X_\zeta | e_\zeta \in i\leadsto \sigma^{2^{k-1}}(i)\}$''.   However, unlike the ``+'' encoding, which requires the 
inclusion of only 1 ``codeword" in each message, the MIN encoding requires the inclusion of 
$O(N \log N)$ i.i.d. ``codewords" in each message in order
to ensure sufficient estimation accuracy~\cite{ModianoShahZussman2006DistSched}.

\section{More Simulation Results}\label{app:more-sim}

\begin{table*}[!thbp]
  \centering
  \setlength{\tabcolsep}{2pt}
  \caption{Average per-port message complexities of SERENADE (bytes).}\label{tab:message-comp}
    \begin{tabular}{@{}l*{16}{r}@{}}
    \toprule
    Traffic patterns &\phantom{ab}& \multicolumn{3}{c}{Uniform} & \phantom{ab}& \multicolumn{3}{c}{Quasi-diagonal}& \phantom{ab} & \multicolumn{3}{c}{Log-diagonal} & \phantom{ab}& \multicolumn{3}{c}{Diagonal} \\
    
    \cmidrule(r){1-1} \cmidrule(l){3-5}\cmidrule(l){7-9}\cmidrule(l){11-13}\cmidrule(l){15-17}
         N   && 64   & 128  & 256 & & 64   & 128  & 256 & & 64   & 128  & 256 & & 64   & 128  & 256 \\
         \midrule
    light ($\rho=0.1$) & & 29.79 & 40.04 & 51.29 && 25.86 & 36.2 & 47.7 && 12.47 & 14.32 & 16.23 && 8.14 & 9.03 & 9.98\\
         \midrule
    moderate ($\rho=0.6$)  & & 34.79 & 44.84 & 55.73  && 31.08 & 40.55 & 50.82 && 21.61 & 25.69 & 29.86 && 14.51 & 16.82 & 19.37 \\
         \midrule
    high ($\rho=0.95$)  & & 35.21 & 45.26 & 56.13   && 28.65 & 37.19 & 46.50   && 20.07 & 23.66 & 27.47    && 15.70 & 18.38 & 21.50\\
         \bottomrule
    \end{tabular}%
\end{table*}%

\subsection{Message Complexities}\label{subsubsec:message-comp}\label{subsec:message-comp}

\begin{figure}[!htbp]
\centering
\includegraphics[width=0.5\textwidth]{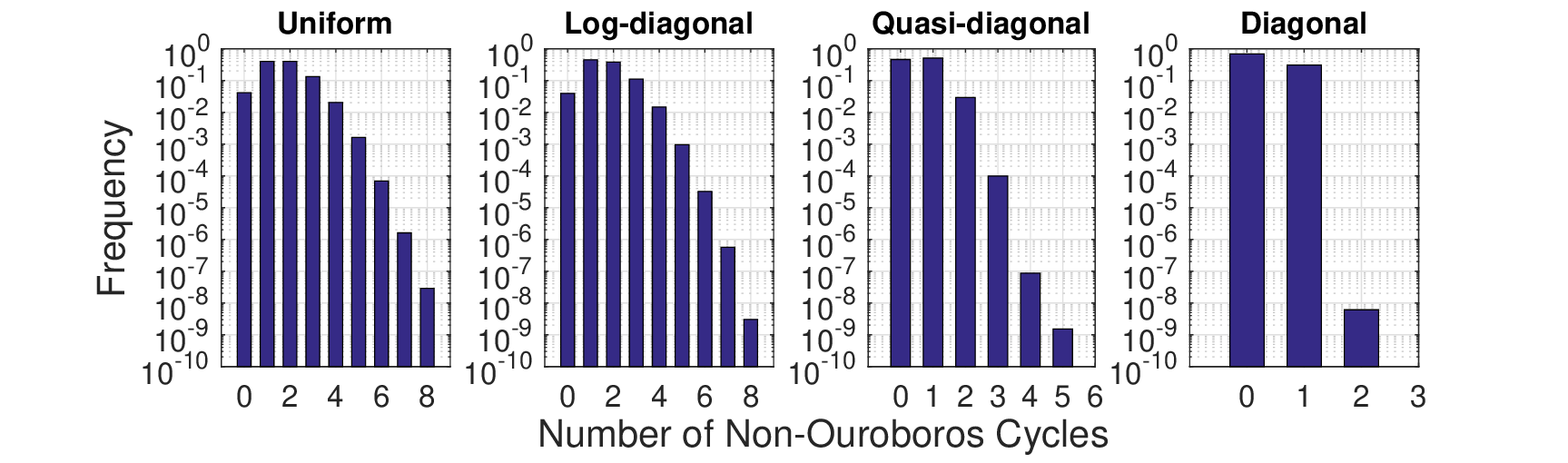}
\caption{Histogram for number of non-ouroboros cycles in SERENADE ($N\!=\!256$, $\rho\!=\!0.6$).}\label{fig:hist-non-o}
\end{figure}

\modifiedHL{
In this section, we investigate the empirical message complexities of SERENADE.   In the interest of space, those of O-SERENADE are not presented 
here, as they are similar. 

\smallskip
\noindent
{\bf Per-Port Message Complexities.} 
\autoref{tab:message-comp} shows the numerical results of the average per-port message complexities (in bytes) of 
SERENADE for $N\!=\!$ 64, 128, and 256 for the $4$ traffic load matrices described above under low, moderate, and 
high offered loads. As explained in~\autoref{subsec: complexity-common}, it suffices to only include 
$w_r(\cdot)\!-\!w_g(\cdot)$, the difference between the red and green weights, in each message. 
This difference can be encoded in $15$ bits (with a single ``sign'' bit), as we assume that each weight can fit in
$14$ bits ({\it i.e.,} no more than 16,384 packets). 
The worst-cast message complexities 
of SERENADE, described in~\autoref{subsec: complexity-common} and~\autoref{subsubsec:comp-bs}, are
$44.25, 53.25, 63$ bytes per port for $N\!=\!$ 64, 128, and 
256 respectively. Comparing them against those values in~\autoref{tab:message-comp}, we can see 
that the average message complexities (per port), under all load factors or traffic patterns, are lower than the worst cases. 
}

\modifiedHL{
\medskip
\noindent
{\bf Number of Non-Ouroboros Cycles.}  
We also measure the number of non-ouroboros cycles in each time slot in SERENADE, 
of which the histograms (on a {\it log scale} along the y-axis) for $N\!=\!$ 256, 
and an offered load of $\rho\!=\!0.6$ under the four different 
load matrices 
are shown in~\autoref{fig:hist-non-o}. The average numbers of non-ouroboros cycles are 
$1.69, 1.61, 0.57, 0.31$ under the uniform, quasi-diagonal, log-diagonal, and diagonal 
load matrices respectively. It is not hard to check that under any of the four different 
load matrices, in more than $99\%$ of instances, there are no more than $4$ non-ouroboros 
cycles per time slot. 
}

\subsection{Delay versus Number of Ports}
\label{subsec: dleay-vs-port}

\begin{figure*}
  \centering
  \subfigure[Under the offered load of 0.6.\label{subfig:moderate-bs}]{
    \centering
    \includegraphics[width=0.88\textwidth]{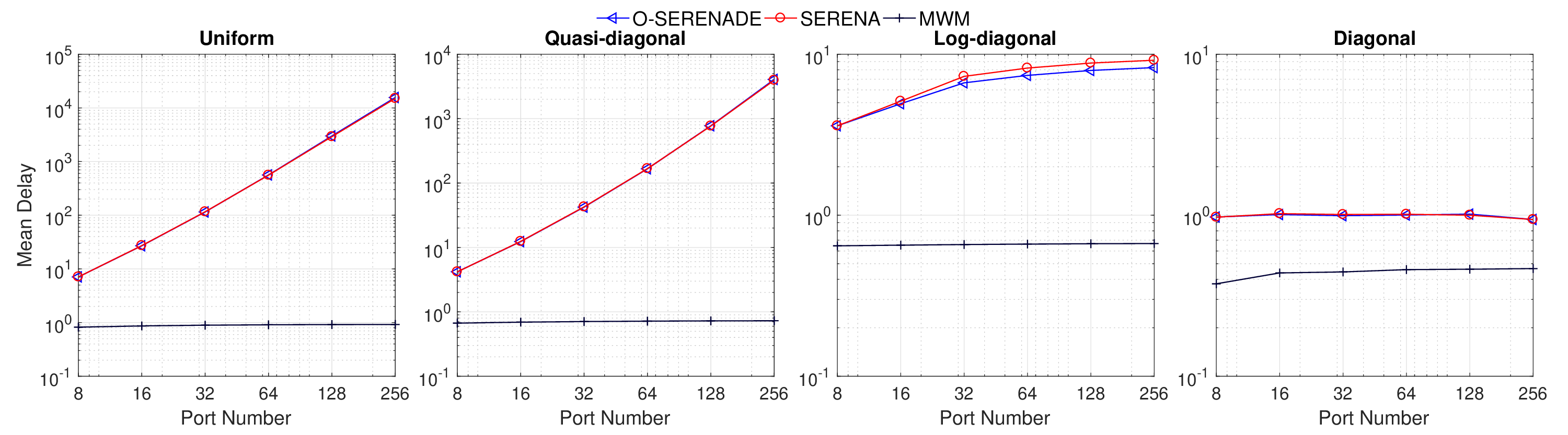}
  }\\
  \subfigure[Under the offered load of 0.95.\label{subfig:heavey-bs}]{
    \centering
    \includegraphics[width=0.88\textwidth]{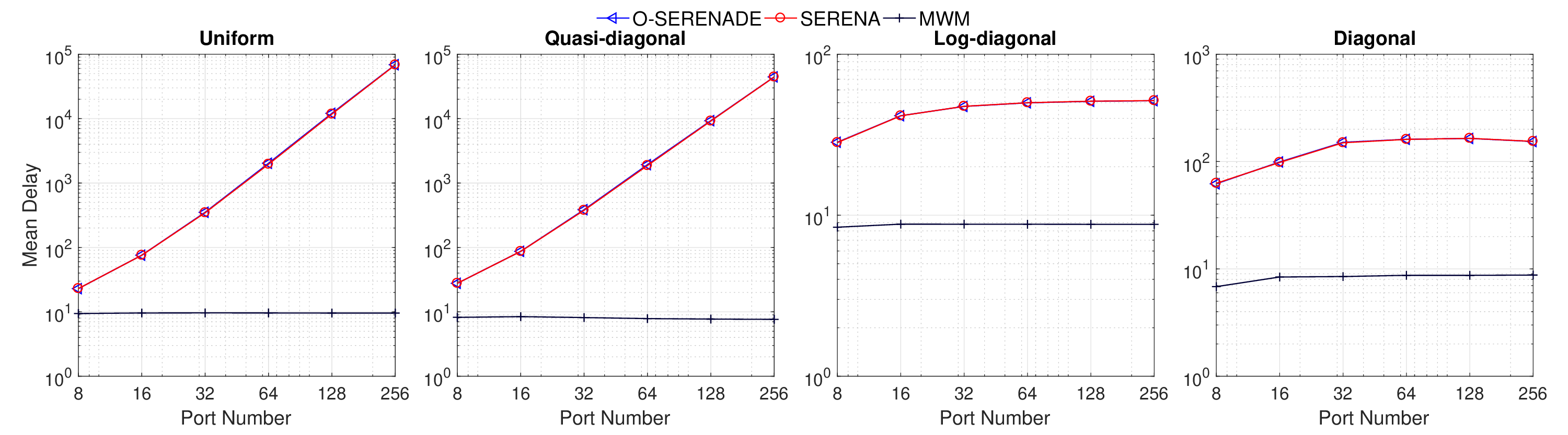}
  }
  \caption{Mean delays v.s. port numbers under
  i.i.d.~Bernoulli traffic arrivals with the $4$ traffic load matrices.}
  \label{fig: mean-delay-vs-port-number}
\end{figure*} 

\modifiedHL{
In this section, we investigate how the mean delay of O-SERENADE 
scales with the number of input/output ports $N$ under {\it i.i.d.} Bernoulli traffic. 
We have simulated the following different $N$ values: $N\!=\!8, 16, 32, 64, 128, 256$.
\autoref{fig: mean-delay-vs-port-number} compares the mean delays for O-SERENADE against 
that for SERENA 
under the $4$ different traffic load matrices with a 
moderate offered load of $0.6$ and a heavy offered load of $0.95$. 
As a benchmark, we also show those of MWM. 
It shows that the scaling behaviors of O-SERENADE are almost the same as that of 
SERENA in terms of mean delays, for all values of $N$. 
More precisely, 
under the log-diagonal and the diagonal load matrices, O-SERENADE
achieves near-optimal scaling ({\it i.e.,} 
nearly constant independent of $N$) of mean delay, whereas under the
uniform and quasi-diagonal load matrices, 
the mean delay grows roughly quadratically with $N$ ({\it i.e.},
$O(N^2)$ scaling).
}

%














\end{document}